\documentclass[12pt]{article}

\usepackage{newtxtext}
\usepackage{amsmath}  

\usepackage{graphicx}

\usepackage[letterpaper,margin=1in]{geometry}

\renewenvironment{abstract}
	{\quotation}
	{\endquotation}

\date{}

\makeatletter
\renewcommand{\fnum@figure}{\textbf{Figure \thefigure}}
\renewcommand{\fnum@table}{\textbf{Table \thetable}}
\makeatother

\usepackage{scicite}

\usepackage{url}

\def\scititle{Generative artificial intelligence reduces social welfare through model collapse}
\title{\bfseries \boldmath \scititle}

\newtheorem{theorem}{Theorem}

\author{
	Fabian Baumann,
	Erol Ak\c{c}ay$^{\dagger}$,
	Joshua B. Plotkin$^{\dagger, \ast}$\\
	University of Pennsylvania\and
    \small$^\dagger$Equal contributions.\\
	\small$^\ast$Corresponding author. Email: jplotkin@sas.upenn.edu
}

\begin{document} 

\maketitle

\begin{abstract} \bfseries \boldmath
\noindent Generative artificial intelligence (genAI) is rapidly reshaping how knowledge and culture are produced and consumed.
Yet generative models are vulnerable to model collapse: when trained on data generated by earlier versions of themselves, their outputs can lose diversity and accuracy. 
This creates a social dilemma, because delegating tasks to genAI can be individually beneficial in the short term even as widespread adoption degrades future model performance. Here we develop a parsimonious model of behavior in collaborative interactions in which individuals can either exert human effort, rely on genAI, or refrain from work altogether. The welfare consequences of genAI are organized by a simple two-dimensional taxonomy: 
the strength of the incentive to perform the task without AI, and the severity of model collapse.
Within this framework, the introduction of genAI---while initially beneficial at the individual level---will reduce social welfare for the most important types of tasks. In addition, habit formation around genAI use can couple otherwise separate domains, so that adoption in low-stakes tasks spills over into high-value tasks and amplifies welfare losses. Together, these results identify a general pathway by which, in the absence of intervention, individually rational adoption of genAI will assuredly and profoundly reduce collective welfare.
\end{abstract}

\noindent Generative artificial intelligence (genAI) is rapidly reshaping how knowledge and culture are produced and consumed \cite{noy2023experimental,assael2025contextualizing,doi:10.1126/science.adz9311,zhang2025deep,rao2025multimodal,swanson2024generative,hao_artificial_2026,gao2024quantifying,epstein2023art,brinkmann2023machine,clark2025extending,farrell2025large,wang2023scientific}, as documented by its swift adoption across domains \cite{bick2026rapid,doi:10.1126/science.adz9311,appel2025anthropic}. Since the launch of ChatGPT in late 2022, large language models (LLMs) have entered mainstream use, and genAI output now permeates the digital ecosystem---including synthetic text, images, music, and code \cite{brooks2024rise,russell2025ai,liang2025quantifying}. This shift has prompted both hopes and fears about the societal consequences of genAI \cite{cui2026effects,becker2025measuring,peng2023impact,dell2023navigating,lin2025persuading,messeri2024artificial,costello2024durably,cheng2026sycophantic}. 

The growing literature on the benefits \cite{tessler2024ai,cui2026effects,becker2025measuring,peng2023impact,dell2023navigating,baum2023fear,peter2025benefits,binz2025should} and risks \cite{kobis2025delegation,baum2023fear,brinkmann2023machine,doshi2024generative,peter2025benefits,binz2025should,kalluri2025computer,haas2026roadmap} to society largely treats genAI as a static technology, assuming performance remains stable as adoption increases. However, generative models are known to be vulnerable to model collapse: their performance can deteriorate when they are retrained on data generated by earlier versions of themselves \cite{shumailov_ai_2024}. 
As models are retrained on their own outputs, the tails of the original distribution of human content disappear, reducing diversity and degrading model performance \cite{seddik2024bad,shumailov_ai_2024,dohmatob2024tale}. This process is already underway: AI-generated content is now common in cultural archives including Wikipedia \cite{brooks2024rise}, news articles \cite{russell2025ai}, and scientific papers \cite{liang2025quantifying}, ensuring that future models will be trained on mixed human–AI corpora. The severity of collapse in a given domain depends on how strongly AI-generated content feeds back into future training.

Model collapse creates a social dilemma: empirical studies have found that genAI usage is associated with substantial productivity gains across a range of tasks \cite{noy2023experimental,cui2026effects}, making adoption individually rational---yet widespread adoption might degrade genAI performance and, ultimately, collective welfare.
To study these feedback dynamics, we develop a mathematical model of a cultural feedback loop in which genAI performance shapes individuals’ propensity to use it, while AI-generated artifacts in turn degrade future performance. Our analysis starts from the recognition that cultural artifacts---and the benefits derived from them---are the result of social interactions. This includes students co-authoring a class assignment, scientists collaborating on a research project, and co-workers jointly producing reports and presentations. Even when a cultural product is seemingly the work of a single individual, such as an employee creating a presentation for their boss, the benefits almost invariably arise from others individuals who interact with it.

We develop the ``collaboration game,'' in which each of $N$ individuals in a population repeatedly chooses among three strategies (Figure~\ref{fig:cartoonish-overview-model}a): perform human work (H), perform no work (N), or delegate their share of the work to genAI (AI). Individuals are paired to collaborate on a joint product (Figure~\ref{fig:cartoonish-overview-model}b), and the total benefit is shared equally between them. Human work and AI-assisted work incur costs $c_h$ and $c_a$ and contribute $b/2$ and $a/2$ to the total benefit, respectively; performing no work incurs no cost and contributes nothing. Over time, individuals adapt their strategies to maximize their own payoff given the strategies played by others in the population.

\subsection*{Model}
\begin{figure}       
\centering
\includegraphics[width=\textwidth]{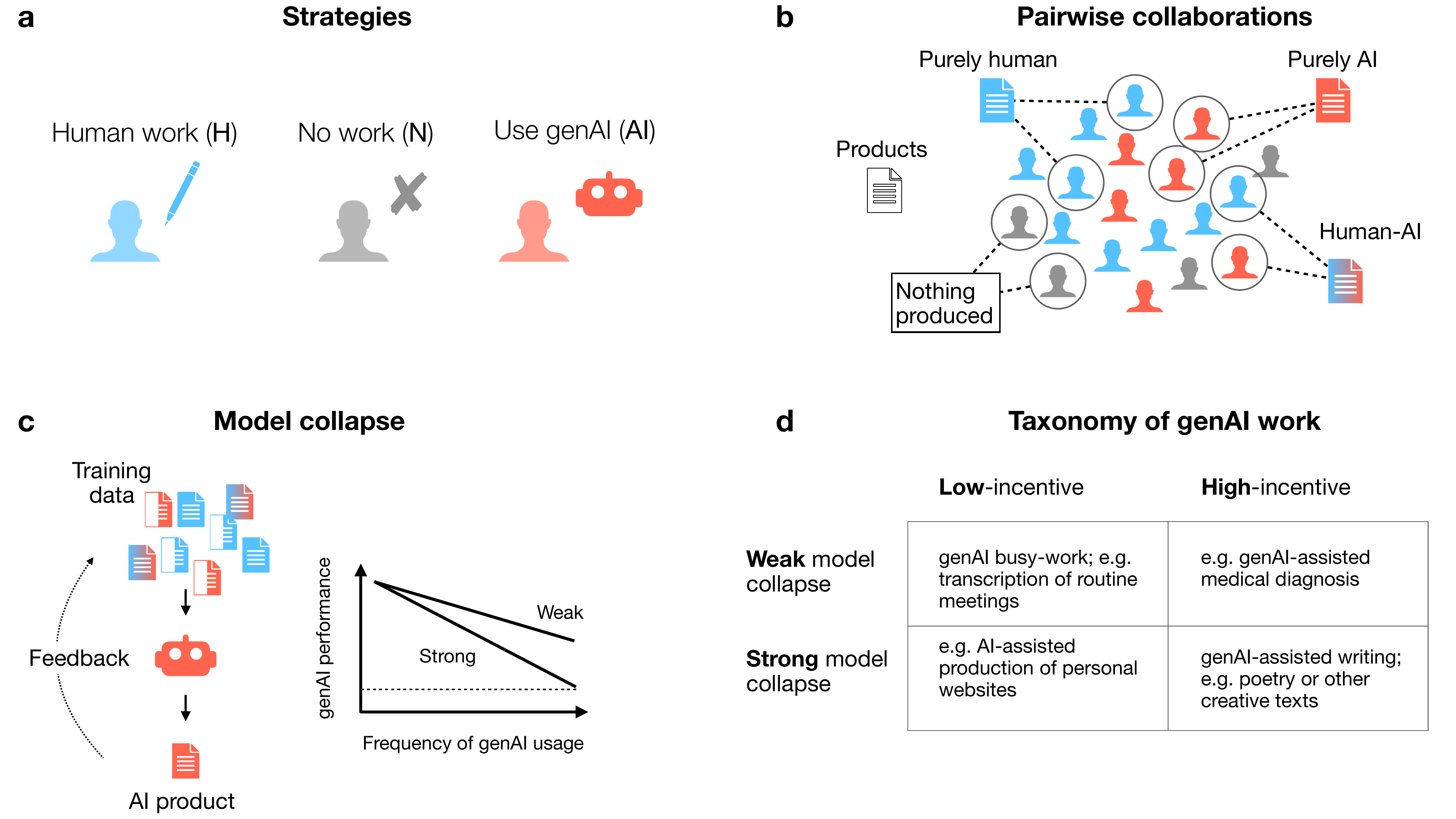}
\caption{\textbf{Model overview.} In the ``collaboration game'' pairs of individuals interact to produce and derive benefits from work products, which enter the cultural domain. Panel~(a) depicts the strategies individuals can choose: perform human work (H), no work (N), or use AI to perform the work (AI). Individuals engage in pairwise interactions to produce a collaborative product, where they incur individual costs $c_i$---depending on their strategies $i\in\{\text{H}, \text{AI}\}$---and share the benefit $b$ of the product (Panel b). We model the collapse of AI performance such that the benefit produced by the AI (AI performance) depends on the overall frequency of AI usage in the population: the more individuals use AI the worse its performance. Model collapse can be weak or strong (c). 
Panel~(d) shows examples of AI work classified along a two-dimensional taxonomy: whether the task is high- or low-incentive, and whether model collapse is weak or strong.
}
  \label{fig:cartoonish-overview-model}
\end{figure}
The  structure of the collaboration game resembles that of a public goods game, formalized by the following payoff matrix
\begin{align}\label{payoffmatrix}
\small
\Pi_\text{AI}=
\bordermatrix{%
& \text{H} & \text{N} & \text{AI}  \cr
\text{H} & b-c_h & \frac{1}{2}b-c_h  & \frac{1}{2}(b+a)-c_h   \cr
\text{N} &  \frac{1}{2}b& 0 & \frac{1}{2}a \cr
\text{AI} & \frac{1}{2}(b+a)-c_a  & \frac{1}{2}a-c_a  &a -c_a \cr
}\,.
\end{align}
Individuals dynamically update their strategies by imitating more successful peers \cite{boyd1988culture}. 
In the limit of a large population ($N\rightarrow\infty$), the evolution of strategies can be approximated by the following system of ordinary differential equations \cite{taylor1978evolutionary}
\begin{equation} 
\dot{x_s} = x_s(\pi_s(\mathbf{x})-\phi(\mathbf{x}))\,, 
\end{equation} 
where $x_s$ denotes the frequency of strategy $s\in\{\text{H}, \text{N}, \text{AI}\}$, and $\mathbf{x}$ is the vector of strategy frequencies with $\sum_s x_s = 1$. 
The average payoff received by an individual using strategy $s$ is given by $\pi_s(\mathbf{x})$, and $\phi(\mathbf{x})$ denotes the average payoff in the population (see \cite{methods} for details). We refer to the population average payoff $\phi(\mathbf{x})$ as \textit{social welfare}.

We assume that the cost of performing human work exceeds the cost of performing genAI work ($c_h>c_a$), whereas the benefit of human work is at least as large as the benefit produced by genAI ($b \geq a$).
To incorporate the phenomenon of AI model collapse we assume that the benefit of using the AI strategy, $a$, depends on the current frequency of individuals who use AI, i.e. $a=a(x_\text{AI})$.
For simplicity of presentation, we assume the benefit of AI-assisted work declines linearly with the frequency of AI usage:
$a(x_{\text{AI}})= b - m x_{\text{AI}}$, as shown in Fig.~\ref{fig:cartoonish-overview-model}c; although our results hold much more generally (see below).
Here $b$ is the benefit produced by human work, and $m>0$ quantifies the strength of model collapse. Larger $m$ implies a faster deterioration of AI performance as AI becomes more prevalent.
These assumptions capture a setting in which AI reduces the private cost of contributing to a collaboration ($c_a<c_h$), while the quality of AI-assisted output declines with population-level AI usage, due to model collapse.

We investigate the impact of AI on social welfare by comparing the long-term outcome in the population against a two-strategy baseline that excludes the possibility of using AI---the interaction simplifies to a two-strategy public goods game with only the strategies H and N available. 
The payoff structure of the reduced game corresponds to the upper-left $2 \times 2$ submatrix of $\Pi_\text{AI}$ in Eq.~\eqref{payoffmatrix} (see \cite{methods} for an explicit definition).

\subsection*{Results}
\begin{figure}       
\centering
\includegraphics[width=0.95\textwidth]{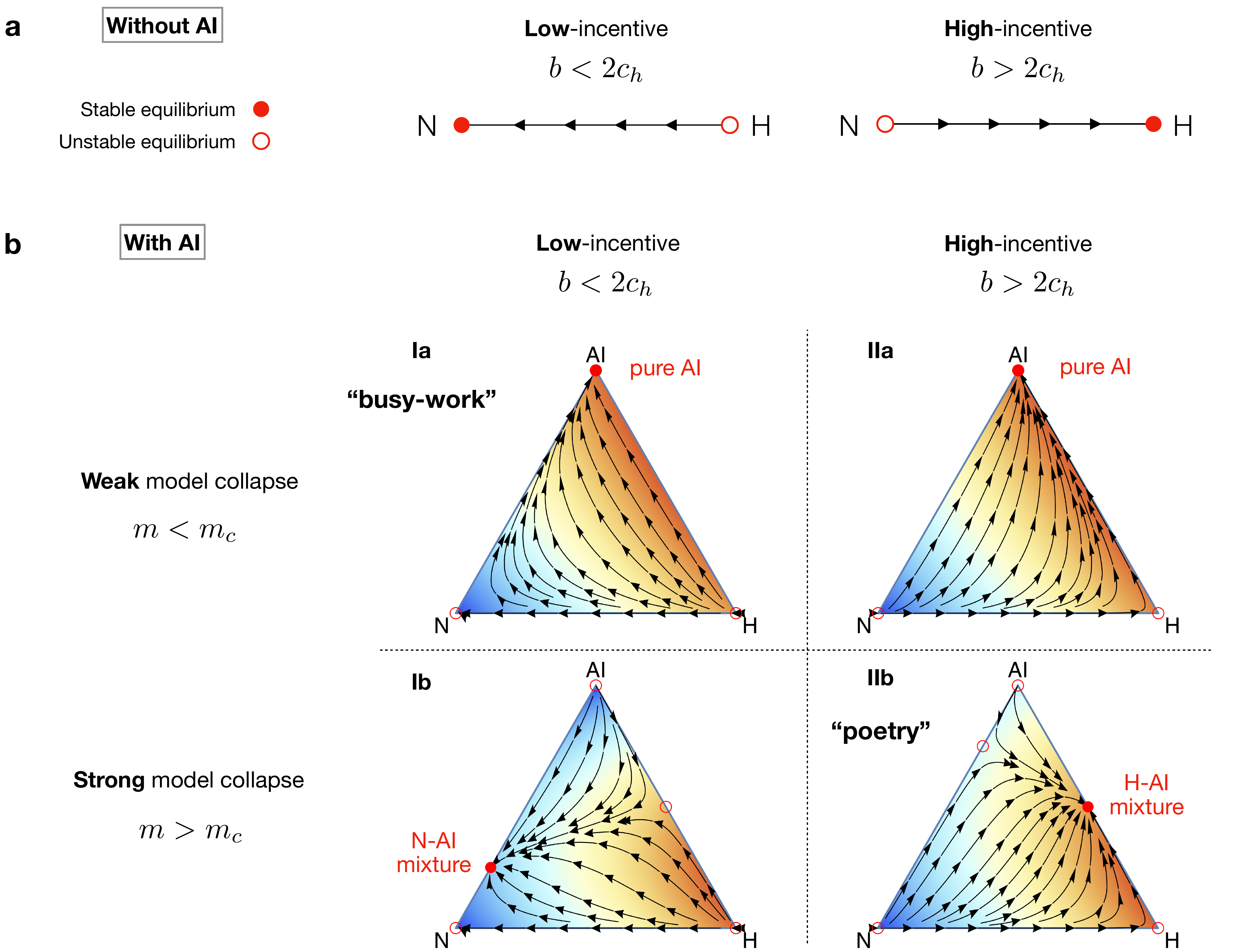}
\caption{\textbf{Strategy evolution in the collaboration game with and without AI}. (a) Without the option to use AI, the collaboration game reduces to a two-strategy public goods game, where individuals can either perform human work (H), or perform no work (N). There are two possible outcomes: if the task is low-incentive ($b<2c_h$) nobody will work (N), otherwise (i.e., for $b>2c_h$, high-incentive), everybody will perform human work (H).
Filled circles indicate stable equilibria, and open circles indicate unstable equilibria. 
(b) In the collaboration game with AI, we classify long-term outcomes using a two-dimensional taxonomy of genAI work: (i) low or high incentive (I/II), and (ii) weak or strong model collapse (a/b).
Games of type~I lead to an all-AI state for weak model collapse (Ia), or to mixed N-AI equilibrium for strong model collapse (Ib). In contrast, games of type~II  lead, again, to an all AI state for weak model collapse (IIa), and a mixed H--AI equilibrium for strong model collapse (IIb). Colors in the ternary plots represent the average payoff (social welfare), increasing from blue to red. 
Model parameters are set to $c_h = 2$, $c_a = 1.5$, $m=0.4$ ($m=2$) for weak (strong) model collapse, and $b=4.5$ ($b=3.5$) for high-incentive (low-incentive) tasks.
}
  \label{fig:dynamics-model-collapse}
\end{figure}

\paragraph{Incentives to work without AI.} The baseline version of the collaboration game---without any option to use genAI---simply tracks what fraction of the population $x_{\text{H}}$ does human work, versus no work at all.  The only possible long-term outcome is either everyone performing human work, or nobody performing any work (Fig.~\ref{fig:dynamics-model-collapse}a). Which of these two outcomes occurs is determined by comparing the shared benefit of effort, $b$, to the personal cost of human work, $c_h$. 
For human work to be stably performed in a society, its benefit $b$ must be greater than $2c_h$, which gives an individual enough net benefit to overcome the incentive to free-ride in the collaboration game (i.e., use strategy N).
We call this the \emph{high-incentive} regime: when $b>2c_h$ the task gets done even without the availability of genAI. Conversely, if $b<2c_h$, the task will not get done in a society without genAI as individuals will choose not to work at all;
we call this the \emph{low-incentive} regime.

\paragraph{Taxonomy of AI-assisted work.}
Introducing genAI produces evolutionary dynamics with three long-run outcomes: a population in which everyone uses genAI, a mixture of individuals contributing either genAI-assisted work or human work, and a mixture of individuals contributing either genAI-assisted work or no work (Fig.~\ref{fig:dynamics-model-collapse}b). For any given parameter set, exactly one of these equilibria is stable.

Which long-run outcome obtains is governed by a two-dimensional taxonomy on domains of genAI usage (Fig.~\ref{fig:dynamics-model-collapse}b). 
The first dimension captures the baseline incentive to do human work as described above, distinguishing high-incentive tasks ($b>2c_h$) from low-incentive tasks ($b<2c_h$). The second dimension captures the strength of model collapse, $m$. Specifically, a critical strength of model collapse, $m_c=\min(b,2c_h)-2c_a$, separates tasks subject to weak model collapse ($m<m_c$) from tasks subject to strong collapse ($m>m_c$).

For tasks subject to weak model collapse ($m<m_c$), the population always converges to the all-AI equilibrium, regardless of whether the task is high- or low-incentive (task types~Ia and~IIa).
Intuitively, genAI continues to perform well even when it is used frequently, when collapse is weak, and so individuals always prefer to rely on AI.  

For tasks subject to strong model collapse ($m>m_c$), the long-term fate of the population depends on the incentive strength. For low-incentive tasks (type~Ib), the population converges to a mixed N--AI equilibrium: some individuals use AI, while the rest free-ride (N). For high-incentive tasks (type~IIb), the population converges instead to a mixed H--AI equilibrium: some individuals contribute human work, while others rely on AI.

The two-by-two taxonomy of tasks corresponds to familiar real-world settings. 
For low-incentive tasks, genAI can enable work that people rarely perform on their own, such as automated transcription or routine summaries of recorded conversations \cite{chen_meetscript_2023}.
Such outputs are often private and do not re-enter training data, corresponding to weak collapse (type~Ia). 
Other low-incentive tasks produce output that is public and likely to be reintegrated into future models---for example, 
automated code documentation in software development \cite{dvivedi2024comparative} or AI-assisted production of personal websites---corresponding to low-incentive tasks subject to strong collapse (type~Ib).
Conversely, some high-incentive domains might be less prone to collapse (type~IIa), because AI serves as a support tool for human decision-making rather than a generator of public output.
AI-assisted medical diagnosis is one such example \cite{gondocs_ai_2024}, though this may change as AI-generated content increasingly enters electronic health records \cite{mccoy2024large}.
Finally, high-incentive domains subject to strong model collapse (type~IIb) are those in which genAI produces culturally salient artifacts that feed back into future training data, such as AI-assisted poetry \cite{kobis_artificial_2021,porter_ai-generated_2024,hitsuwari_does_2023}, fiction writing \cite{yuan2022wordcraft}, scientific research \cite{liang2025quantifying}, and journalism \cite{molla2025artificial}.

We refer to low-incentive tasks subject to weak model collapse (type~Ia) as ``busy-work'' (for presentation purposes): humans would abstain from expending effort absent AI, and AI remains effective even at high adoption. 
A representative example is automated transcription, where the task is adequately solved by AI and further training on the resulting output is limited or not consequential.
Likewise, we refer to high-incentive tasks subject to strong model collapse (type~IIb) as ``poetry'': humans would invest effort absent AI, but retraining on AI-generated output can substantially erode model performance. Representative examples include creative work such as screenplay writing or journalistic prose \cite{appiah2025screenwriter}, where repeated exposure to AI-generated text can yield increasingly stagnant outputs that fail to track the evolving cultural context.

\paragraph{Impact of genAI on social welfare.} 
How does the introduction of genAI affect individual behavior and, in turn, social welfare across the  regimes of collaborative work identified above? To address this question, we compare equilibrium social welfare with and without genAI across the four domains of genAI work (Fig.~\ref{fig:dynamics-model-collapse}b). For each domain, we compute the welfare difference $\Delta S$ at the stable equilibria with versus without the option to use AI. Positive values of $\Delta S$ indicate welfare gains from introducing AI, and negative values indicate welfare losses.

Without any model collapse ($m = 0$), the introduction of AI is either welfare neutral or welfare promoting, with the magnitude of the welfare gain $\Delta S$ determined by the cost of AI use ($c_a$). 
For high-incentive tasks, 
AI replaces human work, and social welfare increases by $\Delta S = c_h - c_a>0$. For low-incentive tasks, 
AI can replace the no-work strategy provided that AI work is net beneficial ($2c_a < b$), and social welfare increases by $\Delta S = b - c_a>0$.
Instead, if AI work is not net beneficial ($b < 2c_a$), the no-work strategy N continues to be an equilibrium after the introduction of AI, and there is no change in social welfare ($\Delta S = 0$).
It is intuitive that the change in social welfare due to AI must be positive  
(or zero) 
when AI does not suffer from model collapse. 
The benefit arises either from a cost reduction when AI replaces human work ($c_a < c_h$), or from the net gain of doing work at all, $b - c_a$, when AI replaces no work.

\begin{figure}       
\centering
\includegraphics[width=\textwidth]{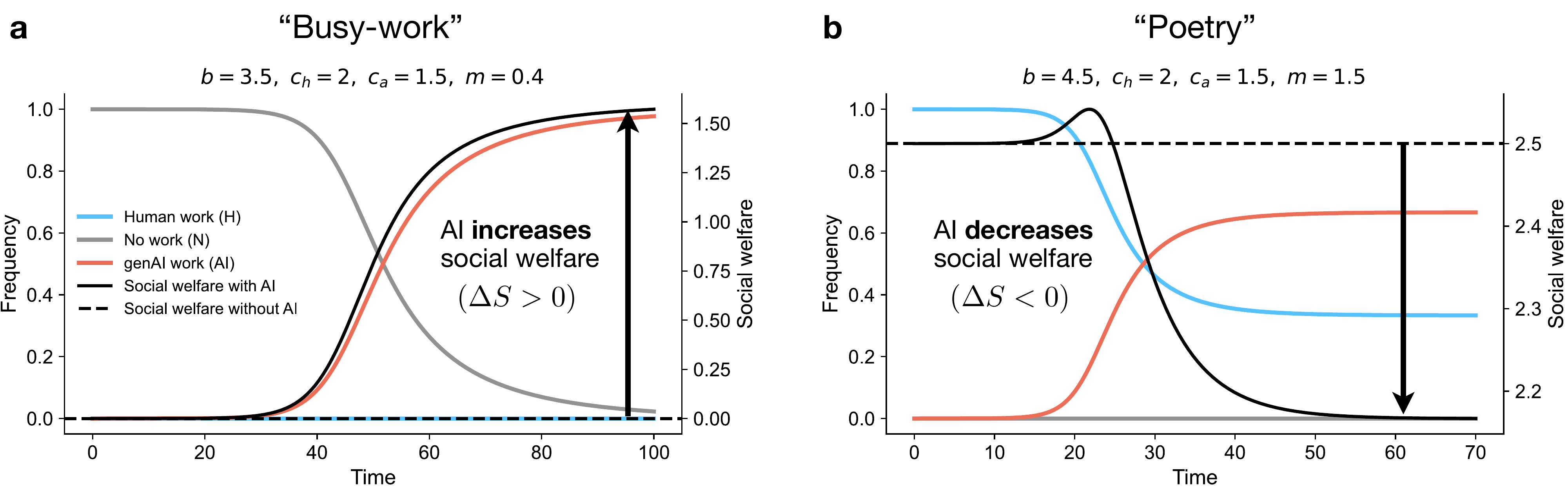}
\caption{\textbf{Impact of AI on social welfare}. Evolution of strategies (H, N, and AI) in the domain of ``busy-work'' (a) and ``poetry'' (b) and the corresponding social welfare in a world with AI. The black dashed lines correspond to social welfare obtained when there is no option to use AI. The  dynamics are initialized at the stable equilibrium obtained without AI, namely all-N for the ``busy-work'' domain and all-H for the ``poetry'' domain.}
  \label{fig:trajectories-welfare}
\end{figure}

Whenever there is some degree of model collapse ($m>0$),
the effect of genAI on welfare is more nuanced. 
For low-incentive tasks,
the introduction of AI is again welfare improving, however, the extent to which it improves social welfare depends on how much work is actually performed by genAI, i.e., how much of the N strategy is replaced by the AI strategy.
For weak model collapse,
no-work is completely replaced by the AI strategy, and $\Delta S=b-m-c_a>0$, so that the introduction of genAI is welfare promoting. Figure~\ref{fig:trajectories-welfare}a shows the evolution of strategies in this domain of ``busy-work'' starting from the equilibrium assumed by the population in the absence of AI. The takeover of AI---i.e., the full replacement of the N-strategy by the AI-strategy---leads to increased social welfare ($\Delta S>0$).
Similarly, for strong model collapse,
where the AI strategy is only partly in the mix, AI is again welfare-promoting ($\Delta S=(b-2c_a)c_a/m>0$). 
Hence, for low-incentive tasks, even those subject to strong model collapse, genAI leads to an increase in social welfare.

The situation changes drastically for high-incentive tasks.
Under weak model collapse, AI fully replaces human work, which can have either a positive or negative impact on social welfare. Specifically, we find $\Delta S = c_h - c_a - m$, which means that genAI is socially beneficial only if the cost savings of using genAI ($c_h-c_a$) exceed the performance loss from model collapse ($m$); otherwise, the introduction of AI is detrimental.
Finally, in the high-incentive regime with strong model collapse (type IIb), the implications of introducing genAI for social welfare are particularly concerning, as formalized in the following theorem (see \cite{methods} for proof):

\begin{theorem}
If a domain of work is high-incentive and there is strong model collapse, such that the population evolves towards a stable mixture of human effort and AI usage, the introduction of genAI always decreases social welfare, i.e., $\Delta S < 0$.
\end{theorem}

This result is particularly pernicious because it applies precisely to those tasks where individuals have a strong incentive to contribute, yet where the introduction of genAI will inevitably lead to over-reliance on genAI despite strong model collapse. The effect of genAI on collective welfare is always negative in this regime.
Practically what follows from Theorem~1 is that if we want to introduce AI to assist humans in some collaborative activity, but predict that some people will still be incentivized to do the task themselves at equilibrium, then we will all end up worse off than having never introduced genAI at all. Although AI enthusiasts may promise that there will always be room for human creativity \cite{epstein2023art} or that human creativity ``becomes more powerful when paired with intelligent machines'' \cite{wind2025creativity}, model collapse means that for any job worth retaining some human activity, society would be better off without genAI at all.

Figure~\ref{fig:trajectories-welfare}b depicts the evolution of strategies in the ``poetry'' domain (type~IIb), where this theorem applies. Starting from the equilibrium state in the absence of genAI, the introduction of genAI causes social welfare initially to increase and then subsequently to decrease below its initial value, causing a long-term loss of welfare in the population.

\paragraph{AI spillover between tasks.} 
So far, we studied the impact of genAI on social welfare in a single task, which depends critically on the incentive strength and the severity of model collapse, i.e., the degree to which genAI performance deteriorates as more individuals adopt it.
Real life, of course, consists of a mix of different tasks and individuals divide their time across multiple types of collaborative tasks. An optimistic scenario for the welfare effects of genAI is when it is only used for tasks where it is welfare-improving, and not for tasks where it is not.

However, this optimistic scenario is constrained by the fact that humans are creatures of habit: repeated use of AI \textit{across} different tasks can lead to habit formation, where individuals tend to rely on AI even in contexts where they usually would not have used AI.
Such habits can couple otherwise independent domains of activity and generate \textit{spillover effects}, in which the use of AI in one domain affects welfare outcomes in another.

\begin{figure}       
  \centering
  \includegraphics[width=\textwidth]{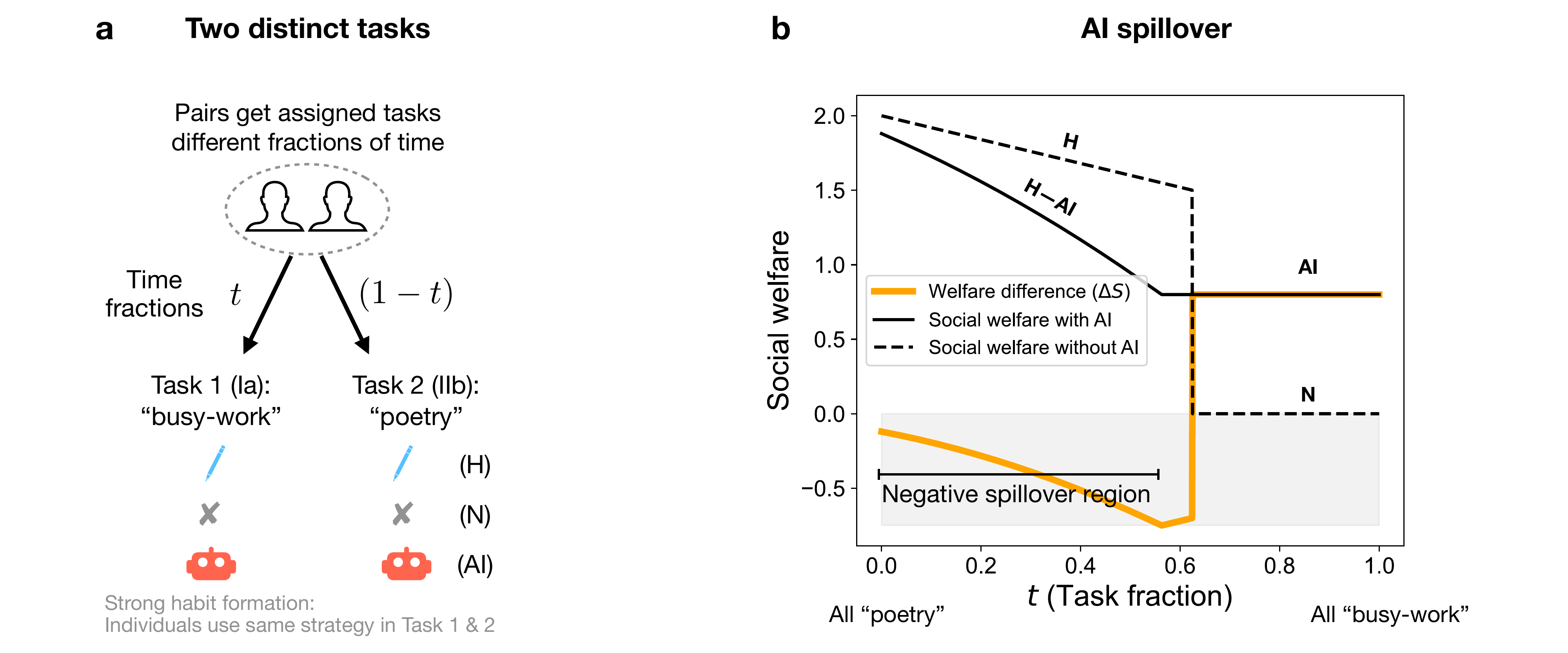}
  \caption{
\textbf{Social welfare and AI spillover.} 
(a) We analyze a setting where individuals are engaged in two different tasks: Task~1 of type~Ia (``busy-work'') for a fraction $t$ of the time and Task~2 of type~IIb (``poetry'') for the remaining fraction $1-t$. 
To capture strong habit formation in the context of genAI, we assume that individuals are constrained to adopt the \textit{same} strategy (H, N, or AI) in both tasks. (b) The resulting equilibrium social welfare for settings with and without AI, and the corresponding difference in welfare, as a function of the proportion time $t$ spend on ``busy-work''.
For $t=0$, this setup reduces to the model with a single task of type IIb, where the welfare difference is \textit{always} negative. On top of that strong habit formation can further amplify negative outcomes through detrimental spillover effects: more individuals switch to using AI when they spend part of their time on ``busy-work'', and habit constraints force them to also use AI in Task~2 of type IIb, where it remains detrimental.
Here, the welfare difference becomes increasingly negative even though larger $t$ values correspond to settings with more ``busy-work'' where AI is beneficial in the single-task case. We call this region of decreasing welfare difference the negative spillover region. The model parameters are $b_1=b_2=3$, $m_1=m_2=1.5$, $c_{a_1}=c_{a_2}=0.7$, $c_{h_1}=1.8$, and $c_{h_2}=1$.}
  \label{fig:two-games}
\end{figure}

To capture this more realistic scenario, we now study a setting in which individuals are engaged in two collaborative tasks from two distinct domains that capture the opposite ends of the welfare effects of genAI: Task~1 is of type~Ia (``busy-work'') where genAI is always welfare improving, while Task~2 is of type~IIb (``poetry'') where it is always welfare reducing (see \cite{methods} for details). 
Pairs of individuals are assigned to one task for a portion of time $0<t<1$, and the other task for the remaining portion $(1-t)$ (Fig.~\ref{fig:two-games}a). 
This setup allows us to analyze how the introduction of genAI in a low-incentive domain with weak collapse can spill over into behavior in a high-incentive domain with strong collapse.
To model the extreme case of strong habit formation, we assume that individuals are restricted to the use of the same strategy in both tasks, as depicted in the bottom part of Fig.~\ref{fig:two-games}a. (In \cite{methods}, we analyze weaker habit formation, with qualitatively similar results.)
We then quantify the welfare implications of introducing genAI in this two-task setting, as a function of the proportion of time $t$ individuals are assigned to each task. 

As a baseline, we first consider strategic outcomes when the population has no recourse to genAI. 
As expected, the population uniformly adopts pure human effort (H) when it spends most of its time engaged in the high-incentive task (type IIb, or ``poetry'').
The population switches to uniformly choose no work (N) when the amount of time $t$ devoted to the low-incentive task (type Ia, or ``busy-work'') increases beyond a critical threshold ($t>t_c$, see \cite{methods} for derivation).
Before this transition, welfare $S$ decreases linearly as the proportion of busy-work increases; and the welfare drops abruptly to $S=0$ for $t>t_c$ as the population stops working. 
These dynamics are depicted as dashed black line in Fig.~\ref{fig:two-games}b.

By contrast, if genAI is available, the population transitions from a mixture of H--AI when most of the time is spent on Task~2 (``poetry''), to uniform adoption of pure AI when most of the time is spent on Task~1 (``busy-work''), as depicted by the solid black line in Fig.~\ref{fig:two-games}b. 
This transition occurs when the proportion of busy-work reaches a critical value $t_l$, which is always smaller than $t_c$ ($t_l < t_c$). On the interval $t \in [0, t_l]$, where the population converges to a stable mixture of human and AI work (H--AI), social welfare decreases quadratically and then becomes constant for $t > t_l$, where the population performs exclusively genAI work (see \cite{methods} for details).

These results show how even if genAI is unambiguously welfare improving for some tasks, its habitual use across tasks can generate a ``spillover'' effect with negative overall welfare effects. The option to use genAI decreases welfare over a large regime (orange line in Fig.~\ref{fig:two-games}b) and the net detrimental effects can increase in magnitude as more time is spent doing ``busy-work"---even though busy-work in isolation would produce societal benefits from genAI usage. 

Spillover arises from the interaction of genAI adoption and habit constraint: as the demand $t$ for ``busy-work'' (Ia) increases, more individuals switch to using genAI, because genAI is helpful for ``busy-work''; but habit constraints force them to also use genAI in Task~2 (of type IIb, ``poetry''), where AI remains detrimental. 
Consequently, the welfare difference becomes increasingly negative, even though larger values of $t$ correspond to settings with more ``busy-work,'' where AI is beneficial in the single-task case. 

\subsection*{Discussion}

The social-welfare impact of genAI hinges on two features of the work domain.  
First, the \emph{incentive strength}---whether the task would be performed even in the absence of genAI.
Second, the \emph{severity of model collapse}---how sharply  performance deteriorates with increasing use and retraining on AI-generated output. These two dimensions jointly determine the long-run outcome. 
For low-incentive tasks, genAI increases welfare by enabling production that would not otherwise occur, even in the presence of model collapse.
But for high-incentive tasks subject to strong model collapse, introducing genAI leads to a stable mixture of human and AI production and a reduction in social welfare relative to a world without genAI.

The dependence on the domain of work clarifies seemingly contradictory narratives about genAI \cite{vaccaro2024combinations}. 
For low-incentive tasks (i.e., ``busy-work''), genAI can increase social welfare by enabling work that would otherwise not occur, because humans would abstain from investing effort.
And for high-incentive tasks, genAI may reduce social welfare.
Most importantly, in regimes where incentives are high and model collapse is strong, detrimental H--AI equilibria arise because AI remains privately attractive despite declining performance.
This dynamic prevents return to purely human work. Practically, this implies that a stable mixture of human and AI production is not evidence of healthy complementarity, but a signature of welfare-reducing feedback.

Spillover between tasks amplifies these concerns. Harm can spread across domains through strong habit formation around use of genAI  \cite{reiter2025student,biswas2024impact,zhai2024effects}. Routine AI use in low-stakes tasks, where it is clearly beneficial, can increase overall reliance; and that reliance can carry over into domains where AI is detrimental. Hence, increasing time spent on ``busy-work'' can, paradoxically, deepen overall welfare losses in ``poetry-like'' domains, even though genAI would be beneficial for the busy-work domain considered in isolation. Spillover also complicates policies for AI safety: evaluating deployment domain by domain will be misleading. 
  
These results complement previous game-theoretic accounts of AI and social-welfare by introducing endogenous performance degradation of AI performance, driven by population behavior. For example, a hiring algorithm used across firms can reduce social welfare despite offering increased individual-level accuracy because it results in firms competing for the same exact candidates \cite{kleinberg2021algorithmic}, similar to the Braess paradox \cite{braess1968paradoxon}, where adding an apparently beneficial resource can worsen collective outcomes. The Braess paradox also appears in models of competition between genAI- and human-based knowledge platforms, such as StackOverflow
\cite{taitler2025braess,taitler2025selective}, where genAI can initially be beneficial yet ultimately reduce welfare by undermining participation in the human platform. Generative AI and agentic agents can also erode incentives to invest human effort in collaborations \cite{taitler2025collaborating} or in the production knowledge that is collectively useful \cite{acemoglu2026ai}. Model collapse reveals a distinct pathway whereby incentives to use AI and feedback on the technology itself can interact to reduce social welfare.

Empirical studies of model collapse \cite{shumailov_ai_2024,guo2023curious,dohmatob2024strong} likely understate its true risk, not only to genAI performance but also to collective welfare. The trajectory of this degradation is difficult to quantify, because early effects may be subtle yet compound into substantial long-term losses. What is already evident, however, is that AI-generated artifacts are being rapidly integrated into important cultural archives---from Wikipedia to news articles to scientific papers \cite{brooks2024rise,russell2025ai,liang2025quantifying} and even spoken language \cite{yakura2024empirical}. Our account of the incentives and feedback this dynamic produces cautions against overly optimistic predictions for genAI’s societal impact. 

From a policy perspective, the key point of intervention is the data feedback loop itself --- namely, limiting untracked reintegration of AI-generated artifacts into future training corpora. Two straightforward approaches are (i) deploying reliable detection and auditing pipelines to identify AI-generated content at scale, and (ii) implementing data-provenance and labeling practices that preserve information about content origin throughout collection, indexing, and training. Both of these remedies require policy, because they may add substantial cost to developing and operating genAI systems. 
It is also important to preserve incentives for human effort in high-value domains---for example, by requiring human-authored components and maintaining skill development in settings where humans produce artifacts not only for economic output but also for pedagogy or cultural value. Finally, organizations and policymakers must anticipate cross-domain coupling and spillover whenever AI tools foster habit formation.
Although our model is intentionally stylized, its qualitative implication is robust: when present-day AI use degrades future performance, individually rational adoption can undermine collective welfare precisely in the domains where human contribution is most valuable.


\clearpage 

%
\bibliography{science_template} 

@article{gondocs_ai_2024,
	title = {{AI} in medical diagnosis: {AI} prediction \& human judgment},
	volume = {149},
	shorttitle = {{AI} in medical diagnosis},
	doi = {10.1016/j.artmed.2024.102769},
	journal = {Artificial Intelligence in Medicine},
	author = {Göndöcs, Dóra and Dörfler, Viktor},
	year = {2024},
	pages = {102769},
}

@article{chen_meetscript_2023,
	title = {{MeetScript}: {Designing} {Transcript}-based {Interactions} to {Support} {Active} {Participation} in {Group} {Video} {Meetings}},
	volume = {7},
	shorttitle = {{MeetScript}},
	doi = {10.1145/3610196},
	number = {CSCW2},
	journal = {Proceedings of the ACM on Human-Computer Interaction},
	author = {Chen, Xinyue and Li, Shuo and Liu, Shipeng and Fowler, Robin and Wang, Xu},
	year = {2023},
	pages = {1--32},
}

@article{shumailov_ai_2024,
	title = {{AI} models collapse when trained on recursively generated data},
	volume = {631},
	doi = {10.1038/s41586-024-07566-y},
	language = {en},
	number = {8022},
	journal = {Nature},
	author = {Shumailov, Ilia and Shumaylov, Zakhar and Zhao, Yiren and Papernot, Nicolas and Anderson, Ross and Gal, Yarin},
	year = {2024},
	pages = {755--759},
}

@article{dohmatob2024strong,
  title={Strong model collapse},
  author={Dohmatob, Elvis and Feng, Yunzhen and Subramonian, Arjun and Kempe, Julia},
  journal={arXiv:2410.04840},
  year={2024}
}

@article{molla2025artificial,
  title={Artificial intelligence and journalism: A systematic bibliometric and thematic analysis of global research},
  author={Molla, Mohammad Al Masum and Ahsan, Md Manjurul},
  journal={Computers in Human Behavior Reports},
  pages={100830},
  year={2025},
  publisher={Elsevier}
}

@article{kobis_artificial_2021,
	title = {Artificial intelligence versus {Maya} {Angelou}: {Experimental} evidence that people cannot differentiate {AI}-generated from human-written poetry},
	volume = {114},
	shorttitle = {Artificial intelligence versus {Maya} {Angelou}},
	doi = {10.1016/j.chb.2020.106553},
	journal = {Computers in Human Behavior},
	author = {Köbis, Nils and Mossink, Luca D.},
	year = {2021},
	pages = {106553},
}

@article{porter_ai-generated_2024,
	title = {{AI}-generated poetry is indistinguishable from human-written poetry and is rated more favorably},
	volume = {14},
	doi = {10.1038/s41598-024-76900-1},
	language = {en},
	number = {1},
	journal = {Scientific Reports},
	author = {Porter, Brian and Machery, Edouard},
	year = {2024},
	pages = {26133},
}

@article{hitsuwari_does_2023,
	title = {Does human–{AI} collaboration lead to more creative art? {Aesthetic} evaluation of human-made and {AI}-generated haiku poetry},
	volume = {139},
	shorttitle = {Does human–{AI} collaboration lead to more creative art?},
	doi = {10.1016/j.chb.2022.107502},
	journal = {Computers in Human Behavior},
	author = {Hitsuwari, Jimpei and Ueda, Yoshiyuki and Yun, Woojin and Nomura, Michio},
	year = {2023},
	pages = {107502},
}

@article{taylor1978evolutionary,
  title={Evolutionary stable strategies and game dynamics},
  author={Taylor, Peter D and Jonker, Leo B},
  journal={Mathematical biosciences},
  volume={40},
  number={1-2},
  pages={145--156},
  year={1978},
  publisher={Elsevier}
}

@article{guo2023curious,
  title={The curious decline of linguistic diversity: Training language models on synthetic text},
  author={Guo, Yanzhu and Shang, Guokan and Vazirgiannis, Michalis and Clavel, Chlo{\'e}},
  journal={arXiv:2311.09807},
  year={2023}
}

@article{doshi2024generative,
	title = {Generative {AI} enhances individual creativity but reduces the collective diversity of novel content},
	volume = {10},
	doi = {10.1126/sciadv.adn5290},
	number = {28},
	journal = {Science Advances},
	author = {Doshi, Anil R. and Hauser, Oliver P.},
	year = {2024},
	pages = {eadn5290},
}

@article{epstein2023art,
  title={Art and the science of generative AI},
  author={Epstein, Ziv and Hertzmann, Aaron and Investigators of Human Creativity and Akten, Memo and Farid, Hany and Fjeld, Jessica and Frank, Morgan R and Groh, Matthew and Herman, Laura and Leach, Neil and others},
  journal={Science},
  volume={380},
  number={6650},
  pages={1110--1111},
  year={2023},
  publisher={American Association for the Advancement of Science}
}

@article{becker2025measuring,
  title={Measuring the impact of early-2025 ai on experienced open-source developer productivity},
  author={Becker, Joel and Rush, Nate and Barnes, Elizabeth and Rein, David},
  journal={arXiv:2507.09089},
  year={2025}
}

@article{farrell2025large,
  title={Large AI models are cultural and social technologies},
  author={Farrell, Henry and Gopnik, Alison and Shalizi, Cosma and Evans, James},
  journal={Science},
  volume={387},
  number={6739},
  pages={1153--1156},
  year={2025},
  publisher={American Association for the Advancement of Science}
}

@article{taitler2025collaborating,
  title={Collaborating with GenAI: Incentives and Replacements},
  author={Taitler, Boaz and Ben-Porat, Omer},
  journal={arXiv:2508.20213},
  year={2025}
}

@article{taitler2025selective,
  title={Selective response strategies for genai},
  author={Taitler, Boaz and Ben-Porat, Omer},
  journal={arXiv:2502.00729},
  year={2025}
}

@article{reiter2025student,
  title={Student (Mis) Use of Generative AI Tools for University-Related Tasks},
  author={Reiter, Leonhard and J{\"o}rling, Moritz and Fuchs, Christoph and {Working group ‘Artificial Intelligence in Higher Education’} and B{\"o}hm, Robert},
  journal={International Journal of Human--Computer Interaction},
  pages={1--14},
  year={2025},
  publisher={Taylor \& Francis}
}

@inproceedings{taitler2025braess,
  title={Braess’s paradox of generative ai},
  author={Taitler, Boaz and Ben-Porat, Omer},
  booktitle={Proceedings of the AAAI Conference on Artificial Intelligence},
  volume={39},
  number={13},
  pages={14139--14147},
  year={2025}
}

@article{kobis2025delegation,
  title={Delegation to artificial intelligence can increase dishonest behaviour},
  author={K{\"o}bis, Nils and Rahwan, Zoe and Rilla, Raluca and Supriyatno, Bramantyo Ibrahim and Bersch, Clara and Ajaj, Tamer and Bonnefon, Jean-Fran{\c{c}}ois and Rahwan, Iyad},
  journal={Nature},
  pages={1--9},
  year={2025},
  publisher={Nature Publishing Group UK London}
}

@article{peng2023impact,
  title={The impact of ai on developer productivity: Evidence from github copilot},
  author={Peng, Sida and Kalliamvakou, Eirini and Cihon, Peter and Demirer, Mert},
  journal={arXiv:2302.06590},
  year={2023}
}

@article{dell2023navigating,
  title={Navigating the jagged technological frontier: Field experimental evidence of the effects of AI on knowledge worker productivity and quality},
  author={Dell'Acqua, Fabrizio and McFowland III, Edward and Mollick, Ethan R and Lifshitz-Assaf, Hila and Kellogg, Katherine and Rajendran, Saran and Krayer, Lisa and Candelon, Fran{\c{c}}ois and Lakhani, Karim R},
  journal={Harvard Business School Technology \& Operations Mgt. Unit Working Paper},
  number={24-013},
  year={2023}
}

@inproceedings{biswas2024impact,
  title={The impact of education level on AI reliance, Habit formation, and usage},
  author={Biswas, Mriganka and Murray, John},
  booktitle={2024 29th International Conference on Automation and Computing (ICAC)},
  pages={1--6},
  year={2024},
  organization={IEEE}
}

@article{zhai2024effects,
  title={The effects of over-reliance on AI dialogue systems on students' cognitive abilities: a systematic review},
  author={Zhai, Chunpeng and Wibowo, Santoso and Li, Lily D},
  journal={Smart Learning Environments},
  volume={11},
  number={1},
  pages={28},
  year={2024},
  publisher={Springer}
}

@article{kleinberg2021algorithmic,
  title={Algorithmic monoculture and social welfare},
  author={Kleinberg, Jon and Raghavan, Manish},
  journal={Proceedings of the National Academy of Sciences},
  volume={118},
  number={22},
  pages={e2018340118},
  year={2021},
  publisher={National Academy of Sciences}
}

@inproceedings{brooks2024rise,
  title={The rise of ai-generated content in wikipedia},
  author={Brooks, Creston and Eggert, Samuel and Peskoff, Denis},
  booktitle={Proceedings of the first workshop on advancing natural language processing for Wikipedia},
  pages={67--79},
  year={2024}
}

@article{yakura2024empirical,
  title={Empirical evidence of Large Language Model's influence on human spoken communication},
  author={Yakura, Hiromu and Lopez-Lopez, Ezequiel and Brinkmann, Levin and Serna, Ignacio and Gupta, Prateek and Soraperra, Ivan and Rahwan, Iyad},
  journal={arXiv:2409.01754},
  year={2024}
}

@techreport{acemoglu2026ai,
  title={AI, Human Cognition and Knowledge Collapse},
  author={Acemoglu, Daron and Kong, Dingwen and Ozdaglar, Asuman},
  year={2026},
  institution={National Bureau of Economic Research}
}

@inproceedings{dvivedi2024comparative,
  title={A comparative analysis of large language models for code documentation generation},
  author={Dvivedi, Shubhang Shekhar and Vijay, Vyshnav and Pujari, Sai Leela Rahul and Lodh, Shoumik and Kumar, Dhruv},
  booktitle={Proceedings of the 1st ACM international conference on AI-powered software},
  pages={65--73},
  year={2024}
}

@article{costello2024durably,
  title={Durably reducing conspiracy beliefs through dialogues with AI},
  author={Costello, Thomas H and Pennycook, Gordon and Rand, David G},
  journal={Science},
  volume={385},
  number={6714},
  pages={eadq1814},
  year={2024},
  publisher={American Association for the Advancement of Science}
}

@article{vaccaro2024combinations,
  title={When combinations of humans and AI are useful: A systematic review and meta-analysis},
  author={Vaccaro, Michelle and Almaatouq, Abdullah and Malone, Thomas},
  journal={Nature Human Behaviour},
  volume={8},
  number={12},
  pages={2293--2303},
  year={2024},
  publisher={Nature Publishing Group UK London}
}

@article{brinkmann2023machine,
  title={Machine culture},
  author={Brinkmann, Levin and Baumann, Fabian and Bonnefon, Jean-Fran{\c{c}}ois and Derex, Maxime and M{\"u}ller, Thomas F and Nussberger, Anne-Marie and Czaplicka, Agnieszka and Acerbi, Alberto and Griffiths, Thomas L and Henrich, Joseph and others},
  journal={Nature Human Behaviour},
  volume={7},
  number={11},
  pages={1855--1868},
  year={2023},
  publisher={Nature Publishing Group UK London}
}

@article{dohmatob2024tale,
  title={A tale of tails: Model collapse as a change of scaling laws},
  author={Dohmatob, Elvis and Feng, Yunzhen and Yang, Pu and Charton, Francois and Kempe, Julia},
  journal={arXiv:2402.07043},
  year={2024}
}

@article{hao_artificial_2026,
	title = {Artificial intelligence tools expand scientists’ impact but contract science’s focus},
	volume = {649},
	doi = {10.1038/s41586-025-09922-y},
	number = {8099},
	journal = {Nature},
	author = {Hao, Qianyue and Xu, Fengli and Li, Yong and Evans, James},
	year = {2026},
	pages = {1237--1243},
}

@book{boyd1988culture,
  title={Culture and the evolutionary process},
  author={Boyd, Robert and Richerson, Peter J},
  year={1988},
  publisher={University of Chicago press}
}

@article{russell2025ai,
  title={AI use in American newspapers is widespread, uneven, and rarely disclosed},
  author={Russell, Jenna and Karpinska, Marzena and Akinode, Destiny and Thai, Katherine and Emi, Bradley and Spero, Max and Iyyer, Mohit},
  journal={arXiv:2510.18774},
  year={2025}
}

@article{braess1968paradoxon,
  title={{\"U}ber ein Paradoxon aus der Verkehrsplanung},
  author={Braess, Dietrich},
  journal={Unternehmensforschung},
  volume={12},
  number={1},
  pages={258--268},
  year={1968},
  publisher={Springer}
}

@article{
doi:10.1126/science.adz9311,
author = {Simone Daniotti  and Johannes Wachs  and Xiangnan Feng  and Frank Neffke },
title = {Who is using AI to code? Global diffusion and impact of generative AI},
journal = {Science},
volume = {0},
number = {0},
pages = {eadz9311},
year = {},
doi = {10.1126/science.adz9311}}

@article{seddik2024bad,
  title={How bad is training on synthetic data? a statistical analysis of language model collapse},
  author={Seddik, Mohamed El Amine and Chen, Suei-Wen and Hayou, Soufiane and Youssef, Pierre and Debbah, Merouane},
  journal={arXiv:2404.05090},
  year={2024}
}

@online{wind2025creativity,
  author  = {Wind, Jerry and Pandya, Mukul and Yao, Deborah},
  title   = {Creativity in the Age of AI},
  year    = {2025},
  month   = oct,
  day     = {20},
  url     = {https://knowledge.wharton.upenn.edu/article/creativity-in-the-age-of-ai/},
  urldate = {2026-02-05},
  note    = {Knowledge at Wharton}
}

@Misc{methods,
  note = {Materials and methods are available as supplementary material},
}

@article{bick2026rapid,
author = {Bick, Alexander and Blandin, Adam and Deming, David J.},
title = {The Rapid Adoption of Generative AI},
journal = {Management Science},
volume = {0},
number = {0},
year = {2026},
doi = {10.1287/mnsc.2025.02523}
}

@article{lin2025persuading,
  title={Persuading voters using human--artificial intelligence dialogues},
  author={Lin, Hause and Czarnek, Gabriela and Lewis, Benjamin and White, Joshua P and Berinsky, Adam J and Costello, Thomas and Pennycook, Gordon and Rand, David G},
  journal={Nature},
  pages={1--8},
  year={2025},
  publisher={Nature Publishing Group UK London}
}

@inproceedings{yuan2022wordcraft,
  title={Wordcraft: story writing with large language models},
  author={Yuan, Ann and Coenen, Andy and Reif, Emily and Ippolito, Daphne},
  booktitle={Proceedings of the 27th International Conference on Intelligent User Interfaces},
  pages={841--852},
  year={2022}
}

@article{liang2025quantifying,
  title={Quantifying large language model usage in scientific papers},
  author={Liang, Weixin and Zhang, Yaohui and Wu, Zhengxuan and Lepp, Haley and Ji, Wenlong and Zhao, Xuandong and Cao, Hancheng and Liu, Sheng and He, Siyu and Huang, Zhi and others},
  journal={Nature Human Behaviour},
  pages={1--11},
  year={2025},
  publisher={Nature Publishing Group UK London}
}

@article{swanson2024generative,
  title={Generative AI for designing and validating easily synthesizable and structurally novel antibiotics},
  author={Swanson, Kyle and Liu, Gary and Catacutan, Denise B and Arnold, Autumn and Zou, James and Stokes, Jonathan M},
  journal={Nature Machine Intelligence},
  volume={6},
  number={3},
  pages={338--353},
  year={2024},
  publisher={Nature Publishing Group UK London}
}

@article{clark2025extending,
  title={Extending minds with generative AI},
  author={Clark, Andy},
  journal={Nature Communications},
  volume={16},
  number={1},
  pages={4627},
  year={2025},
  publisher={Nature Publishing Group UK London}
}

@article{assael2025contextualizing,
  title={Contextualizing ancient texts with generative neural networks},
  author={Assael, Yannis and Sommerschield, Thea and Cooley, Alison and Shillingford, Brendan and Pavlopoulos, John and Suresh, Priyanka and Herms, Bailey and Grayston, Justin and Maynard, Benjamin and Dietrich, Nicholas and others},
  journal={Nature},
  volume={645},
  number={8079},
  pages={141--147},
  year={2025},
  publisher={Nature Publishing Group UK London}
}

@article{rao2025multimodal,
  title={Multimodal generative AI for medical image interpretation},
  author={Rao, Vishwanatha M and Hla, Michael and Moor, Michael and Adithan, Subathra and Kwak, Stephen and Topol, Eric J and Rajpurkar, Pranav},
  journal={Nature},
  volume={639},
  number={8056},
  pages={888--896},
  year={2025},
  publisher={Nature Publishing Group UK London}
}

@article{baum2023fear,
  title={From fear to action: AI governance and opportunities for all},
  author={Baum, Kevin and Bryson, Joanna and Dignum, Frank and Dignum, Virginia and Grobelnik, Marko and Hoos, Holger and Irgens, Morten and Lukowicz, Paul and Muller, Catelijne and Rossi, Francesca and others},
  journal={Frontiers in Computer Science},
  volume={5},
  pages={1210421},
  year={2023},
  publisher={Frontiers Media SA}
}

@article{binz2025should,
  title={How should the advancement of large language models affect the practice of science?},
  author={Binz, Marcel and Alaniz, Stephan and Roskies, Adina and Aczel, Balazs and Bergstrom, Carl T and Allen, Colin and Schad, Daniel and Wulff, Dirk and West, Jevin D and Zhang, Qiong and others},
  journal={Proceedings of the National Academy of Sciences},
  volume={122},
  number={5},
  pages={e2401227121},
  year={2025},
  publisher={National Academy of Sciences}
}

@article{peter2025benefits,
  title={The benefits and dangers of anthropomorphic conversational agents},
  author={Peter, Sandra and Riemer, Kai and West, Jevin D},
  journal={Proceedings of the National Academy of Sciences},
  volume={122},
  number={22},
  pages={e2415898122},
  year={2025},
  publisher={National Academy of Sciences}
}

@article{kalluri2025computer,
  title={Computer-vision research powers surveillance technology},
  author={Kalluri, Pratyusha Ria and Agnew, William and Cheng, Myra and Owens, Kentrell and Soldaini, Luca and Birhane, Abeba},
  journal={Nature},
  volume={643},
  number={8070},
  pages={73--79},
  year={2025},
  publisher={Nature Publishing Group UK London}
}

@article{haas2026roadmap,
title={A roadmap for evaluating moral competence in large language models},
author={Haas, Julia and Bridgers, Sophie and Manzini, Arianna and Henke, Benjamin and May, Joshua and Levine, Sydney and Weidinger, Laura and Shanahan, Murray and Lum, Kristian and Gabriel, Iason and others},
journal={Nature},
volume={650},
number={8102},
pages={565--573},
year={2026},
publisher={Nature Publishing Group UK London}
}

@article{messeri2024artificial,
  title={Artificial intelligence and illusions of understanding in scientific research},
  author={Messeri, Lisa and Crockett, Molly J},
  journal={Nature},
  volume={627},
  number={8002},
  pages={49--58},
  year={2024},
  publisher={Nature Publishing Group UK London}
}

@article{appiah2025screenwriter,
  author  = {Appiah, Kwame Anthony},
  title   = {I'm a Screenwriter. {Is} It All Right if {I} Use {A.I.}?},
  journal = {The New York Times Magazine},
  year    = {2025},
  month   = oct,
  day     = {4},
  note    = {The Ethicist column},
  url     = {https://www.nytimes.com/2025/10/04/magazine/magazine-email/screenwriter-ai-ethics.html}
}

@article{mccoy2024large,
  title={Large language models and the degradation of the medical record},
  author={McCoy, Liam G and Manrai, Arjun K and Rodman, Adam},
  journal={The New England journal of medicine},
  volume={391},
  number={17},
  pages={1561--1564},
  year={2024}
}

@article{gao2024quantifying,
  title={Quantifying the use and potential benefits of artificial intelligence in scientific research},
  author={Gao, Jian and Wang, Dashun},
  journal={Nature Human Behaviour},
  volume={8},
  number={12},
  pages={2281--2292},
  year={2024},
  publisher={Nature Publishing Group UK London}
}

@article{wang2023scientific,
  title={Scientific discovery in the age of artificial intelligence},
  author={Wang, Hanchen and Fu, Tianfan and Du, Yuanqi and Gao, Wenhao and Huang, Kexin and Liu, Ziming and Chandak, Payal and Liu, Shengchao and Van Katwyk, Peter and Deac, Andreea and others},
  journal={Nature},
  volume={620},
  number={7972},
  pages={47--60},
  year={2023},
  publisher={Nature Publishing Group UK London}
}

@article{appel2025anthropic,
  title={Anthropic economic index report: Uneven geographic and enterprise ai adoption},
  author={Appel, Ruth and McCrory, Peter and Tamkin, Alex and McCain, Miles and Neylon, Tyler and Stern, Michael},
  journal={arXiv:2511.15080},
  year={2025}
}

@article{noy2023experimental,
  title={Experimental evidence on the productivity effects of generative artificial intelligence},
  author={Noy, Shakked and Zhang, Whitney},
  journal={Science},
  volume={381},
  number={6654},
  pages={187--192},
  year={2023},
  publisher={American Association for the Advancement of Science}
}

@article{cui2026effects,
	title = {The {Effects} of {Generative} {AI} on {High}-{Skilled} {Work}: {Evidence} from {Three} {Field} {Experiments} with {Software} {Developers}},
	shorttitle = {The {Effects} of {Generative} {AI} on {High}-{Skilled} {Work}},
	journal = {Management Science},
	author = {Cui, Kevin Zheyuan and Demirer, Mert and Jaffe, Sonia and Musolff, Leon and Peng, Sida and Salz, Tobias},
	year = {2026},
}

@article{cheng2026sycophantic,
  title={Sycophantic AI decreases prosocial intentions and promotes dependence},
  author={Cheng, Myra and Lee, Cinoo and Khadpe, Pranav and Yu, Sunny and Han, Dyllan and Jurafsky, Dan},
  journal={Science},
  volume={391},
  number={6792},
  pages={eaec8352},
  year={2026},
  publisher={American Association for the Advancement of Science}
}

@article{zhang2025deep,
  title={Deep generative models design mRNA sequences with enhanced translational capacity and stability},
  author={Zhang, He and Liu, Hailong and Xu, Yushan and Huang, Haoran and Liu, Yiming and Wang, Jia and Qin, Yan and Wang, Haiyan and Ma, Lili and Xun, Zhiyuan and others},
  journal={Science},
  volume={390},
  number={6773},
  pages={eadr8470},
  year={2025},
  publisher={American Association for the Advancement of Science}
}

@article{tessler2024ai,
  title={AI can help humans find common ground in democratic deliberation},
  author={Tessler, Michael Henry and Bakker, Michiel A and Jarrett, Daniel and Sheahan, Hannah and Chadwick, Martin J and Koster, Raphael and Evans, Georgina and Campbell-Gillingham, Lucy and Collins, Tantum and Parkes, David C and others},
  journal={Science},
  volume={386},
  number={6719},
  pages={eadq2852},
  year={2024},
  publisher={American Association for the Advancement of Science}
}
\bibliographystyle{sciencemag}

\newpage


\renewcommand{\thefigure}{S\arabic{figure}}
\renewcommand{\thetable}{S\arabic{table}}
\renewcommand{\theequation}{S\arabic{equation}}
\renewcommand{\thepage}{S\arabic{page}}
\setcounter{figure}{0}
\setcounter{table}{0}
\setcounter{equation}{0}
\setcounter{page}{1} 


\begin{center}
\section*{Supplementary Materials for\\ \scititle}


\author{
	Fabian Baumann,
	Erol Ak\c{c}ay,
	Joshua B. Plotkin\\
    \vspace{0.3cm}
	University of Pennsylvania\\
}
\end{center}


\vspace{1.5cm}
\noindent This supplement provides the formal derivations underlying the results presented in the main text. We begin by reviewing the mathematical foundations of our modeling approach, which is based on tools from evolutionary game theory. 
We specify the collaboration game with and without the option to use genAI, and derive the corresponding equilibria and welfare results. 
We then prove Theorem~1 from the main text, showing that the welfare-loss result holds for any decreasing model-collapse function. 
Finally, we analyze the two-task model, deriving equilibria under varying degrees of habit formation and characterizing the negative spillover effects on social welfare. 

\section*{Materials and Methods}
\subsection*{Modeling framework}
\subsubsection*{Evolution of strategies}

We formulate our model within the framework of evolutionary game theory. Individuals engage in pairwise collaborations structured as a modified public goods game, where the payoff to a focal individual $i$ interacting with a peer $j$ is given by the payoff matrix $\Pi_{ij}$. Individuals update their strategies through payoff-biased imitation, a classical model of behavioral evolution. We consider an infinite, well-mixed population in continuous time, in which individuals are randomly matched to play a symmetric game and strategies spread in proportion to their payoff. Let $\mathbf{x}=(x_1,\dots,x_\mathrm{N})$ denote the population state, where $x_i$ is the frequency of strategy $i$ and $\sum_i x_i=1$.

Given a population state $\mathbf{x}$, the expected payoff to strategy $i$ is
\begin{equation*}
\pi_i(\mathbf{x})=\sum\limits^{n}_{j=1}\Pi_{ij}x_j,
\end{equation*}
and the average population payoff (social welfare) is
\begin{equation*}
\phi(\mathbf{x})=\sum\limits_{i=1}^{n}x_i\pi_i(\mathbf{x})=\mathbf{x}^\top \Pi\mathbf
{x}\,.
\end{equation*}
Under replicator dynamics, the frequency of each strategy evolves according to
\begin{equation*}
\dot{x}_i=x_i(\pi_i(\mathbf{x})-\phi(\mathbf{x}))\;\;\;i=1,\dots,n.
\end{equation*}

The social welfare of a given equilibrium $\mathbf{x}^*$ is obtained by evaluating the average payoff at that equilibrium, i.e.,  
\begin{equation*}
\phi(\mathbf{x}^*) = {\mathbf{x}^*}^\top \Pi\,\mathbf{x}^*.
\end{equation*}

\subsubsection*{The collaboration game}
\textit{Collaborations with the option to use genAI.---}The collaboration game with the option to use genAI is defined by the payoff matrix
\begin{align*}
\Pi_\text{AI}=
\bordermatrix{%
& \text{H} & \text{N} & \text{AI}  \cr
\text{H} & -c_h +b & -c_h + \frac{b}{2} & -c_h +\frac{[b+a(\mathbf{x})]}{2}  \cr
\text{N} & \frac{b}{2} & 0 & \frac{a(\mathbf{x})}{2} \cr
\text{AI} & -c_a +\frac{[b+a(\mathbf{x})]}{2} & -c_a +\frac{a(\mathbf{x})}{2} & -c_a+a(\mathbf{x}) \cr
}\,,
\end{align*}
in which individuals choose among three strategies: engage in human work (H), refrain from work (N), or use genAI to perform the work (AI). Strategies H and AI incur costs $c_h$ and $c_a$, respectively, while strategy N incurs no cost and contributes no benefit. Human work produces a benefit $b$, and AI-assisted work produces a benefit $a(\mathbf{x})$, where the function $a(\mathbf{x})$ captures the dependence of genAI performance on the population's strategy profile. For most of the analysis below, we model genAI-collapse with a linear function. That is, we assume that $a$ decreases linearly with the frequency of AI usage: $a(x_\text{AI})=b-mx_\text{AI}$. However, our main result (Theorem 1 in the main text) is true for general model collapse functions, as explained below.

In the main text we restrict attention to the case where human effort is at least as beneficial as AI-assisted work ($b \geq a$), human effort is more costly than AI effort ($c_h > c_a$), and AI-assisted work is always net beneficial ($a > 2c_a$). These assumptions select a subset of possible outcomes. Here we characterize the full set of equilibria without imposing these restrictions.

Substituting $x_\text{AI}=1-x_\text{H}-x_\text{N}$ reduces the dynamics to a system of two coupled replicator equations,
\begin{align*}
    \dot{x}_\text{H}&=x_\text{H}(\pi_\text{H}-\phi)\\
    \dot{x}_\text{N}&=x_\text{N}(\pi_\text{N}-\phi)
\end{align*}
with expected payoffs
\begin{align*}
    \pi_\text{H}&=\frac{1}{2}\,\bigl(-2c_h - b(-2+x_\text{N}) - m(-1+x_\text{H}+x_\text{N})^2\bigr)\\
    \pi_\text{N}&=\frac{1}{2}\,\bigl(b - b x_\text{N} - m(-1+x_\text{H}+x_\text{N})^2\bigr)
\end{align*}
and mean population payoff
\begin{align*}
    \phi&=b - c_h x_\text{H} - b x_\text{N} + c_a(-1 + x_\text{H} + x_\text{N}) - m(-1 + x_\text{H} + x_\text{N})^2\,.
\end{align*}

Without model collapse ($m=0$), the system admits only pure equilibria: the all-H state $\left(x^{*(1)}_\text{H}=1,\, x^{*(1)}_\text{N}=0\right)$, stable if $b>c_h$ and $c_h<c_a$; the all-N state $\left(x^{*(2)}_\text{H}=0,\, x^{*(2)}_\text{N}=1\right)$, stable if $b<2c_h$ and $b<2c_a$; and the all-AI state $\left(x^{*(3)}_\text{H}=0,\, x^{*(3)}_\text{N}=0\right)$, stable if $b>2c_a$ and $c_a<c_h$.

With model collapse ($m>0$), the system admits five equilibria---three pure and two mixed---whose stability is determined by standard Jacobian analysis. The pure equilibria are: the all-H state $\left(x^{*(1)}_\text{H}=1,\, x^{*(1)}_\text{N}=0\right)$, stable if $b>c_h$ and $c_h<c_a$; the all-N state $\left(x^{*(2)}_\text{H}=0,\, x^{*(2)}_\text{N}=1\right)$, stable if $b<2c_h$ and $b<2c_a$; and the all-AI state $\left(x^{*(3)}_\text{H}=0,\, x^{*(3)}_\text{N}=0\right)$, stable if $b>2c_a+m$ and $c_h>c_a+m/2$.

The two mixed equilibria are: a mixed N--AI equilibrium,
\[
\left(x^{*(4)}_\text{H}=0,\; x^{*(4)}_\text{N}=\frac{2c_a-b+m}{m}\right),
\]
stable if $b>2c_a$, $m>b-2c_a$, and $2c_h>b$; and a mixed H--AI equilibrium,
\[
\left(x^{*(5)}_\text{H}=\frac{2c_a-2c_h+m}{m},\; x^{*(5)}_\text{N}=0\right),
\]
stable if $c_h>c_a$, $m>2(c_h-c_a)$, and $b>2c_h$.

To identify the socially optimal (SO) state in the absence of model collapse ($m=0$), we rewrite the mean payoff as
\begin{equation*}
    \phi(m=0)=(1-x_\text{N})\underbrace{(b-c_a)}_{>0}+\underbrace{(c_a-c_h)}_{<0}x_\text{H}\,.
\end{equation*}
This expression is maximized at $(x^\text{SO}_\text{H}=0,\, x^\text{SO}_\text{N}=0)$, so the pure-AI state is socially optimal when $m=0$. Although the pure-AI state can also arise as an evolutionary outcome for $m>0$, it is socially optimal only when $m=0$.

Because the public goods structure penalizes free-riding, the social optimum generically lies on the H--AI edge of the simplex (i.e., $x_\text{N}=0$). Solving $\frac{\partial}{\partial x_\text{H}}\phi\big|_{x_\text{N}=0}=0$ for $m>0$ gives
\begin{equation}\label{eq:social_optimum_m>0}
\left(x^{\text{SO}}_\text{H}=\frac{c_a-c_h+2m}{2m},\;\;\;\; x^{\text{SO}}_\text{AI}=\frac{c_h-c_a}{2m}\right).
\end{equation}
It follows from \eqref{eq:social_optimum_m>0} that neither mixed equilibrium is socially optimal, and that the social optimum always involves more human work than either equilibrium: $x^{\text{SO}}_\text{H}>x^{*(5)}_\text{H}>x^{*(4)}_\text{H}$.

\textit{Collaborations without the option to use genAI---}To quantify the welfare impact of introducing genAI, we compare the outcomes of the full collaboration game (described above) to a two-strategy baseline in which only human work (H) and no work (N) are available. The baseline is defined by the payoff matrix
\begin{align*}
\Pi_\text{no-AI}=
\bordermatrix{%
& \text{H} & \text{N}  \cr
\text{H} & -c_h +b & -c_h + \frac{b}{2}  \cr
\text{N} & \frac{b}{2} & 0  \cr
}\,.
\end{align*}
Since $x_\text{N} = 1 - x_\text{H}$, the dynamics reduce to a single replicator equation,
\begin{equation}\label{eq:two-strategy-without-AI-replicator}
   \dot{x}_\text{H} = \frac{1}{2} (b - 2c_h)(1-x_\text{H})x_\text{H}\,.
\end{equation}
Equation~\eqref{eq:two-strategy-without-AI-replicator} admits two equilibria: the all-H state ($x^{*(1)}_\text{H} = 1$), stable for $b > 2c_h$, and the all-N state ($x^{*(2)}_\text{H} = 0$), stable for $b < 2c_h$. The corresponding social welfare is $\phi(x_\text{H}) = (b-c_h)x_\text{H}$. The population converges to a socially optimal state when $b>2c_h$ or $b<c_h$; for $c_h<b<2c_h$, however, the population settles on the all-N state, which is suboptimal since the net social benefit of universal human work, $b - c_h$, is positive in this range.

\subsection*{Proof of Theorem 1}

We prove that any \textit{stable} mixed H--AI equilibrium of the collaboration game yields strictly lower social welfare than the all-H baseline. The proof holds for a general collapse function $a(x_\text{AI})$ and does not require linearity.

Consider the collaboration game defined by the payoff matrix
\begin{align*}
\Pi_\text{AI}=
\bordermatrix{%
& \text{H} & \text{N} & \text{AI}  \cr
\text{H} & -c_h +b & -c_h + \frac{b}{2} & -c_h +\frac{b+a(\mathbf{x})}{2}  \cr
\text{N} & \frac{b}{2} & 0 & \frac{a(\mathbf{x})}{2} \cr
\text{AI} & -c_a +\frac{b+a(\mathbf{x})}{2} & -c_a +\frac{a(\mathbf{x})}{2} & -c_a+a(\mathbf{x}) \cr
}\,,
\end{align*}
where $a(\mathbf{x})$ is a general function describing AI performance as a function of the population state. This function need not be linear; we only assume that it is continuously differentiable. We assume $c_h > c_a$ (performing human work is more costly than delegating work to AI) and $b > 2c_h$ 
(the task is high-incentive).

\subsubsection*{Mixed strategy equilibrium condition} At a mixed H--AI equilibrium, $x_\text{N}=0$, so the population state reduces to $\mathbf{x}=(1-x_\text{AI},\, 0,\, x_\text{AI})$ with $x_\text{AI}\in(0,1)$. Coexistence of H and AI requires equal expected payoffs, $\pi_\text{AI}(\mathbf{x})=\pi_\text{H}(\mathbf{x})$, which yields
\begin{equation}\label{eq:conditionHAI}
    b-a(x_\text{AI})=2(c_h-c_a)\,.
\end{equation}
This condition is independent of the functional form of $a$, and states that AI performance has declined sufficiently for the cost advantage of AI to be exactly offset by the quality advantage of human work. We use the existence of such a mixed strategy equilibrium as the generalized characterization of strong model collapse.

\textit{Definition 1:} AI exhibits strong model collapse when Eq.~\eqref{eq:conditionHAI} holds for $0<x_\text{AI}<1$. 

Note that for a linear model collapse function, this definition reduces to the one given in the main text. 

\subsubsection*{Stability of the mixed equilibrium} 
We verify that a mixed H--AI equilibrium satisfying Eq.~\eqref{eq:conditionHAI} can be stable under two conditions on $a$.

First, stability along the H--AI edge requires that deviations from the equilibrium are self-correcting. The payoff difference $\pi_\text{AI} - \pi_\text{H}$ must be decreasing in $x_\text{AI}$ at the equilibrium, so that an increase in AI usage makes AI relatively less attractive. This holds whenever
\begin{equation}\label{eq:stability-edge}
    a'(x_\text{AI}^*) < 0\,,
\end{equation}
i.e., AI performance is locally decreasing in AI usage---precisely the model-collapse assumption.

Second, stability against invasion by strategy N requires that free-riding is not profitable at the equilibrium. Evaluating $\pi_\text{N} - \pi_\text{H}$ at $x_\text{N}=0$ gives
\begin{equation}
    \pi_\text{N} - \pi_\text{H} = c_h - \frac{b}{2}\,,
\end{equation}
which is negative whenever $b > 2c_h$---the high-incentive condition.

Hence, any collapse function $a(x_\text{AI})$ that is decreasing at the interior solution of Eq.~\eqref{eq:conditionHAI} produces a stable mixed H--AI equilibrium, provided the task is high-incentive.

\subsubsection*{Welfare comparison} The social welfare at any H--AI equilibrium is
\begin{equation}
    S=b - c_h + \bigl(a(x_\text{AI}) - b + c_h - c_a\bigr)\, x_\text{AI}\,.
\end{equation}
In the absence of genAI, the high-incentive regime ($b>2c_h$) yields the all-H equilibrium with welfare $b-c_h$. The welfare difference is therefore
\begin{equation}
    \Delta S = \bigl(a(x_\text{AI}) - b + c_h - c_a\bigr)\, x_\text{AI}\,.
\end{equation}
Substituting the equilibrium condition~\eqref{eq:conditionHAI}, i.e., $a(x_\text{AI}) - b = -2(c_h - c_a)$, gives
\begin{equation}
    \Delta S = -x_\text{AI}(c_h-c_a)\,.
\end{equation}
Since $c_h>c_a$ and $x_\text{AI}>0$, we conclude $\Delta S<0$. That is, the introduction of genAI strictly reduces social welfare at any stable mixed H--AI equilibrium.

\subsection*{AI spillover between tasks}

Here we present in detail the two-task extension of the model, in which individuals simultaneously engage in two collaborative tasks, Task~1 and Task~2. A time-allocation parameter $t\in(0,1)$ specifies the fractions of time devoted to each task: $t$ to Task~1 and $1-t$ to Task~2.

Each task is an independent public goods game with its own parameter set: $\{b_1, c_{a_1}, c_{h_1}, m_1\}$ for Task~1 and $\{b_2, c_{a_2}, c_{h_2}, m_2\}$ for Task~2. Following the setup in Fig.~4a of the main text, we assign Task~1 to be of type ``\textit{busy-work}'' (type~Ia) and Task~2 to be of type ``\textit{poetry}'' (type~IIb). This is achieved by setting $b_1=b_2=b$, $c_{a_1}=c_{a_2}=c_a$, $m_1=m_2=m$, and $c_{h_1}>c_{h_2}$, together with the conditions $b<2c_{h_1}$ and $m<2c_{h_1}-2c_a$ 
(ensuring Task~1 is low-incentive and subject to weak collapse), as well as $m>2(c_{h_2}-c_a)$ and $b>2c_{h_2}$ 
(ensuring Task~2 is high-incentive and subject to strong collapse). 
A strategy in the two-task model is a tuple XY, where X and Y denote the strategies used in Task~1 and Task~2, respectively.

\subsubsection*{Strong habit formation}
In the main text we present results for the limiting case of strong habit formation, in which each individual is constrained to use the \textit{same} strategy in both tasks. The effective strategy set therefore reduces to $s\in\{\text{H}, \text{N}, \text{AI}\}$, where each label stands for the corresponding homogeneous tuple (H,H), (N,N), or (AI,AI).

\textit{Evolution of strategies.---}The payoff matrix of the two-task model under strong habit formation is the time-weighted average of the two single-task payoff matrices,
\begin{equation*}
\tilde{\Pi}_\text{AI}=t\,\Pi_{\text{AI},1}+(1-t)\,\Pi_{\text{AI},2}\,,
\end{equation*}
where the subscripts 1 and 2 refer to the parameter sets of Task~1 and Task~2, respectively.

Substituting $x_\text{AI}=1-(x_\text{H}+x_\text{N})$ yields two coupled replicator equations,
\begin{align*}
    \dot{x}_\text{H}&=\tfrac{1}{2} x_\text{H} \bigl((-2 c_a - 2 c_{h_2} (-1 + t) + 2 c_{h_1} t + m (-1 + x_\text{H})) (-1 + x_\text{H}) + (b - 2 (c_a + m - m x_\text{H})) x_\text{N} + m x_\text{N}^2\bigr)\,,\\
   \dot{x}_\text{N}&= \tfrac{1}{2} x_\text{N} \bigl(2 c_a + m - 2 (c_a + m + c_{h_2} (-1 + t) - c_{h_1} t) x_\text{H} + m x_\text{H}^2 + 
   b (-1 + x_\text{N}) - 2 c_a x_\text{N} + m x_\text{N} (-2 + 2 x_\text{H} + x_\text{N})\bigr)\,.
\end{align*}

These dynamics admit five equilibria, of which exactly two are ever stable. For $0 \leq t < t_l$, the stable equilibrium is a mixed H--AI state,
\begin{align*}
    \left(x_\text{H}^* = \frac{2 c_a + m + 2 c_{h_2} (-1 + t) - 2 c_{h_1} t}{m}\,, \quad
    x_\text{N}^* = 0\,, \quad
    x_\text{AI}^* = -\frac{2 (c_a + c_{h_2} (-1 + t) - c_{h_1} t)}{m}\right)\,,
\end{align*}
where $t_l = \frac{c_a - c_{h_2} + m/2}{c_{h_1} - c_{h_2}}$. For $t_l < t \leq 1$, the mixed equilibrium loses stability and the pure all-AI state,
\begin{align*}
    \left(x_\text{H}^* = 0\,, \quad x_\text{N}^* = 0\,, \quad x_\text{AI}^* = 1\right)\,,
\end{align*}
becomes the unique stable equilibrium.

The change in equilibrium stability at $t = t_l$ gives rise to a piecewise expression for social welfare as a function of $t$:
\begin{align}\label{eq:SwithAIinfinite}
S(t) =
\begin{cases}
    \frac{-2 c_a^2 + b m + \left( c_{h_2} (-1 + t) - c_{h_1} t \right) \left( m - 2 c_{h_2} (-1 + t) + 2 c_{h_1} t \right) + 
4 c_a \left( c_{h_2} + c_{h_1} t - c_{h_2} t \right)
}{m}, & \text{for } t < t_l, \\[6pt]
    b-c_a-m, & \text{for } t > t_l\,.
\end{cases}
\end{align}
For $t < t_l$, the population is in a mixed H--AI equilibrium and welfare decreases quadratically with $t$. For $t > t_l$, the population adopts pure AI in both tasks and welfare is constant in $t$.

\textit{Welfare difference.---}To quantify the impact of introducing genAI on social welfare, we compare the welfare obtained in the two-task model with AI to that of a baseline model in which individuals have no access to the AI strategy. In this baseline, individuals choose the same strategy---either H or N---in both tasks, yielding a combined payoff matrix
\begin{equation*}
    \tilde{\Pi}_\text{no-AI}=t\,\Pi_{\text{no-AI},1}+(1-t)\,\Pi_{\text{no-AI},2}\,,
\end{equation*}
where the indices 1 and 2 refer to the parameter sets of Task~1 and Task~2, respectively. Under the parametrization above, 
Task~1 is low-incentive and Task~2 is high-incentive.

Substituting $x_\text{N} = 1 - x_\text{H}$ reduces the dynamics to a single replicator equation:
\begin{align*}
    \dot{x}_\text{H}=-\frac{1}{2} (b - 2 (c_{h_2} + c_{h_1} t - c_{h_2} t)) (-1 + x_\text{H}) x_\text{H}\,.
\end{align*}
This admits two pure equilibria: the all-H state $(x_\text{H}^* = 1)$ and the all-N state $(x_\text{H}^* = 0)$. Given $c_{h_1} >b> c_{h_2}$, the all-H state is stable for $t < t_c \equiv \frac{b/2 - c_{h_2}}{c_{h_1} - c_{h_2}}$, and the all-N state is stable for $t > t_c$. The corresponding social welfare is
\begin{align}\label{eq:SwithoutAIinfinite}
S(t) =
\begin{cases}
    t(-c_{h_1}+b)+(1-t)(-c_{h_2}+b), & \text{for } t < t_c, \\[6pt]
    0, & \text{for } t > t_c\,,
\end{cases}
\end{align}
which decreases linearly with $t$ for $t < t_c$ and drops to zero once the population ceases to work.

Combining Eqs.~\eqref{eq:SwithAIinfinite} and~\eqref{eq:SwithoutAIinfinite}, the welfare difference due to the introduction of genAI is
\begin{align}\label{eq:deltaS-infinite-switching-costs}
\Delta S =
\begin{cases}
- \dfrac{2 \left( c_a + c_{h_2} (-1 + t) - c_{h_1} t \right)^2}{m}, & \text{for } t < t_l, \\[6pt]
- c_a + c_{h_2} - m + c_{h_1} t - c_{h_2} t, & \text{for } t_l < t < t_c, \\[6pt]
b - c_a - m, & \text{for } t \ge t_c\,.
\end{cases}
\end{align}
The welfare difference is initially negative and decreases quadratically for $t < t_l$, then increases linearly for $t_l < t < t_c$, and remains constant for $t \ge t_c$. This piecewise behavior is depicted in Fig.~4b of the main text.

\subsubsection*{Weak habit formation}
The main text presents results for the limiting case of strong habit formation. Here we show that the same qualitative phenomenon---the emergence of a negative spillover region---persists under weak habit formation. In this setting, individuals may adopt different strategies across tasks, giving rise to all nine possible strategy combinations: HH, HN, HA, NH, NN, NA, AH, AN, and AA, where, for example, HA denotes strategy H in Task~1 and strategy A in Task~2.

We formalize weak habit formation by introducing a finite switching cost $c$, incurred whenever an individual uses different strategies across the two tasks. This cost vanishes for the three homogeneous strategy tuples (HH, NN, AA), which require no behavioral switch.

\begin{figure}       
\centering
\includegraphics[width=\textwidth]{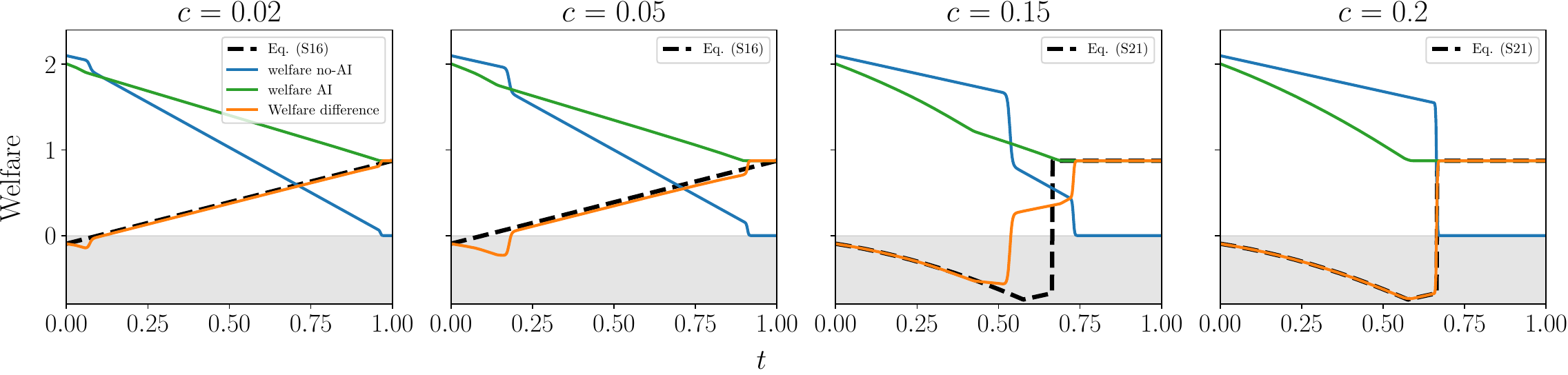}
\caption{\textbf{Finite switching costs for two simultaneous games.} Numerical results for two simultaneous games (Game~1 of type Ia and Game~2 of type IIb) with finite switching costs $c$ depicting social welfare with (green) and without (green) AI and the welfare difference (orange). Dashed lines depict the analytical expressions derived for zero (Eq.~\eqref{eq:deltaS-zero-switching-cost}) and infinite switching costs (Eq.~\eqref{eq:deltaS-infinite-switching-costs}).
The model parameters are identical to those in Fig.~4 of the main text, i.e.,  $b_1=b_2=3$, $m_1=m_2=1.5$, $c_{a_1}=c_{a_2}=0.7$, $c_{h_1}=1.8$, and $c_{h_2}=1$.}
  \label{fig:two-game-phi_vs_T}
\end{figure}

\textit{Evolution of strategies.---}Under weak habit formation, the payoff matrix is no longer a simple weighted average of two single-task matrices. Instead, it is a $9 \times 9$ matrix defined over the full set of strategy combinations. To construct it, we first introduce a generalized version of the single-task $3 \times 3$ payoff matrix:
\begin{align*}
\small
\Pi_{\text{AI},\gamma}(S_i, S_j, \sigma)=
\bordermatrix{%
& \text{H} & \text{N} & \text{AI}  \cr
\text{H} & -c_{h_\gamma} +b_\gamma & -c_{h_\gamma} + \frac{b_\gamma}{2} & -c_{h_\gamma} +\frac{b_\gamma+a_\gamma(\sigma)}{2} \cr
\text{N} & \frac{b_\gamma}{2} & 0 & \frac{a_\gamma(\sigma)}{2} \cr
\text{AI} & -c_{a_\gamma} +\frac{b_\gamma+a_\gamma(\sigma)}{2} & -c_{a_\gamma} +\frac{a_\gamma}{2} & -c_{a_\gamma}+a_\gamma(\sigma) \cr
}\,,
\end{align*}
where $S_i, S_j \in \{H, N, AI\}$ denote the strategies of the focal individual $i$ and their collaborator $j$, respectively. The index $\gamma \in \{1, 2\}$ identifies the task and allows for task-specific parameters, and $\sigma$ denotes the aggregate frequency of strategies that contribute to model collapse in task $\gamma$.

The payoff to a focal individual $i$ employing the tuple strategy $[S_{i,1}, S_{i,2}]$ against a collaborator $j$ using $[S_{j,1}, S_{j,2}]$ is then
\begin{align}\label{eq:general-two-game-matrix-with-AI}
    \tilde{\Pi}_\text{AI}([S_{i,1}, S_{i,2}],[S_{j,1}, S_{j,2}])=t\,&\Pi_{\text{AI},1}(S_{i,1},S_{j,1},\sigma_1)\notag\\+(1-t)\,&\Pi_{\text{AI},2}\left(S_{i,2},S_{j,2},\sigma_2\right) -(1-\delta_{S_{i,1},S_{i,2}})c\,,
\end{align}
where $\delta$ is the Kronecker delta, $c$ is the switching cost incurred when different strategies are used across tasks, and $\sigma_1$, $\sigma_2$ are the aggregate frequencies of AI-using strategies in Task~1 and Task~2, respectively. For instance, strategy AA contributes to model collapse in both tasks, whereas HA contributes only in Task~2.

In summary, the full strategy set comprises nine tuple strategies with associated frequencies: HH ($x_1$), HN ($x_2$), HA ($x_3$), NN ($x_4$), NH ($x_5$), NA ($x_6$), AA ($x_7$), AH ($x_8$), and AN ($x_9$), yielding $\sigma_1 = x_7 + x_8 + x_9$ and $\sigma_2 = x_3 + x_6 + x_7$.

\textit{Welfare difference.---}To quantify the welfare impact of introducing genAI in the two-task model, we compare the outcome with AI to a baseline in which individuals choose among four strategy tuples: HH ($x_1$), HN ($x_2$), NH ($x_3$), and NN ($x_4$). The baseline payoff matrix is
\begin{align}\label{eq:payoff-mat-with-no-ai-finite}
    \tilde{\Pi}_\text{no-AI}([S_{i,1},S_{i,2}],[S_{j,1},S_{j,2}])=t \Pi_{\text{no-AI},1}(S_{i,1},S_{j,1})+(1-t)&\Pi_{\text{no-AI},2}\left(S_{i,2},S_{j,2}\right) -(1-\delta_{S_{i,1},S_{i,2}})c\,,
\end{align}
with
\begin{align}\label{eq:sg-without-AI}
\small
\Pi_{\text{no-AI},\gamma}(S_{1}, S_{2})=
\bordermatrix{%
& \text{H} & \text{N}   \cr
\text{H} & -c_{h_\gamma} +b_\gamma & -c_{h_\gamma} + \frac{b_\gamma}{2}  \cr
\text{N} & \frac{b_\gamma}{2} & 0  \cr
}\,.
\end{align}

The payoff matrices defined in Eqs.~\eqref{eq:general-two-game-matrix-with-AI} and \eqref{eq:payoff-mat-with-no-ai-finite} give rise to nine and four replicator equations, respectively. We integrate both systems numerically until stationarity: for each value of $t \in [0,1]$ and switching cost $c$, the system is initialized on the simplex and evolved until the time derivative of the state vector becomes negligibly small. The resulting stationary frequencies $\mathbf{x}^*$ yield social welfare $\phi(t) = \mathbf{x}^{*\top} \Pi\,\mathbf{x}^*$.

The results are shown in Fig.~\ref{fig:two-game-phi_vs_T}, with panels corresponding to increasing values of the switching cost $c$ from left to right. For very small $c$, the welfare difference $\Delta S$ is well approximated across most values of $t$ by the analytical expression for zero switching cost (Eq.~\eqref{eq:deltaS-zero-switching-cost}), shown as a black dashed line in the two leftmost panels. However, even for $c$ as small as $0.02$, a negative spillover region begins to emerge near $t = 0$. As $c$ increases (e.g., $c = 0.05$), this region expands, although $\Delta S$ remains nearly linear and closely tracks Eq.~\eqref{eq:deltaS-zero-switching-cost} over a broad intermediate range of $t$. The pattern changes qualitatively for larger switching costs (e.g., $c = 0.15$): the near-linear regime breaks down entirely, and $\Delta S$ increasingly resembles the highly nonlinear behavior predicted by the strong-habit-formation limit (Eq.~\eqref{eq:deltaS-infinite-switching-costs}; black dashed line in the two rightmost panels). By $c = 0.2$, the numerical solution and the analytical prediction for strong habit formation have effectively converged.

\subsubsection*{No habit formation}
The case of no habit formation corresponds to zero switching costs ($c=0$). Under our parametrization (Task~1 of type~Ia; Task~2 of type~IIb), the dynamics can be analyzed by restricting attention to the four strategies that survive in equilibrium: NA ($x_1$), NH ($x_2$), AH ($x_3$), and AA ($x_4$). The aggregate frequencies of AI usage in each task are then $\sigma_1 = x_3 + x_4$ and $\sigma_2 = x_1 + x_4$.

The payoff to a focal individual employing tuple strategy $[S_{i,1}, S_{i,2}]$ against a peer using $[S_{j,1}, S_{j,2}]$ is
\begin{align*}
    \tilde{\Pi}_\text{AI}([S_{i,1},S_{i,2}],[S_{j,1},S_{j,2}])
    =t\,\Pi_{\text{AI},1}(S_{i,1},S_{j,1},\sigma_1)+(1-t)\,\Pi_{\text{AI},2}(S_{i,2},S_{j,2},\sigma_2)\,,
\end{align*}
where $[S_{i,1}, S_{i,2}] \in \{\text{NA}, \text{NH}, \text{AH}, \text{AA}\}$.

Substituting $x_1=1-(x_2+x_3+x_4)$ yields three replicator equations for the frequencies of NH ($x_2$), AH ($x_3$), and AA ($x_4$):
\begin{multline*}
\dot{x}_{2} = \tfrac{1}{2} x_{2} \big(
    m - m t - 2 m x_{2} + 2 m t x_{2} + m x_{2}^2 - m t x_{2}^2 - 2 m x_{3} - b t x_{3} + 2 m t x_{3}\\ + 2 m x_{2} x_{3} - 2 m t x_{2} x_{3} + m x_{3}^2 - 2 c_{h_2} (-1 + t)(-1 + x_{2} + x_{3})\\ - b t x_{4} + 2 m t x_{3} x_{4} + m t x_{4}^2 + 2 c_a (1 - x_{2} - x_{3} + t (-1 + x_{2} + 2 x_{3} + x_{4}))
\big)
\end{multline*}
\begin{multline*}
\dot{x}_{3} = \tfrac{1}{2} x_{3} \big(m + b t - m t - 2 m x_{2} + 2 m t x_{2} + m x_{2}^2 - m t x_{2}^2 - 2 m x_{3} \\ - 
   b t x_{3} + m t x_{3} + 2 m x_{2} x_{3} - 2 m t x_{2} x_{3} + m x_{3}^2 - 2 c_{h_2} (-1 + t) (-1 + x_{2} + x_{3}) \\- t (b + m - 2 m x_{3}) x_{4} + m t x_{4}^2 + 2 c_a (1 - x_{2} - x_{3} + t (-2 + x_{2} + 2 x_{3} + x_{4}))\big)
\end{multline*}
\begin{multline*}
\dot{x}_{4} = \tfrac{1}{2} x_{4} \big(2 c_{h_2} x_{2} - m x_{2} - 2 c_{h_2} t x_{2} + m t x_{2} + m x_{2}^2 - m t x_{2}^2 + 2 c_{h_2} x_{3} \\-
    m x_{3} - 2 c_{h_2} t x_{3} + 2 m x_{2} x_{3} - 2 m t x_{2} x_{3} + m x_{3}^2 + m t (-1 + 2 x_{3}) x_{4} \\+
    m t x_{4}^2 - b t (-1 + x_{3} + x_{4}) + 2 c_a (-x_{2} - x_{3} + t (-1 + x_{2} + 2 x_{3} + x_{4}))\big)
\end{multline*}

The system has a unique stable equilibrium, which persists for all values of $t \in [0,1]$:
$x_{2}^{*} = 0$, $x_{3}^{*} = \frac{2c_a - 2c_{h_2} + m}{m}$, and 
$x_{4}^{*} = -\frac{2(c_a - c_{h_2})}{m}$.
The corresponding social welfare is
\begin{equation*}
S(t)=\frac{
-2 c_a^2 + 4 c_a c_{h_2} - 2 c_{h_2}^2 + b m - 
c_{h_2} m + \left( c_a - c_{h_2} - m \right) \left( 2 c_a - 2 c_{h_2} + m \right) t
}{m}\,,
\end{equation*}
which is linear in $t$, as shown in Fig.~\ref{fig:two-game-phi_vs_T}.

\textit{Welfare difference.---}In the corresponding baseline without AI, the strategy set consists of the tuples HH ($x_1$), HN ($x_2$), NH ($x_3$), and NN ($x_4$). With zero switching costs, the payoff matrix is
\begin{align*}
    \tilde{\Pi}_\text{no-AI}([S_{i,1},S_{i,2}],[S_{j,1},S_{j,2}])=t\,\Pi_{\text{no-AI},1}(S_{i,1},S_{j,1})+(1-t)\,\Pi_{\text{no-AI},2}(S_{i,2},S_{j,2})\,,
\end{align*}
where $\Pi_{\text{no-AI},\gamma}$ is defined in Eq.~\eqref{eq:sg-without-AI}.

Setting $b_1 = b_2 = b$ with $c_{h_1}>b>c_{h_2}$ and substituting $x_1=1-(x_2+x_3+x_4)$ yields three replicator equations for HN ($x_2$), NH ($x_3$), and NN ($x_4$):
\begin{align*}
  \dot{x}_2 &= \frac{1}{2} x_2 \bigl(2 c_{h_2} (-1 + t) (-1 + x_2 + x_4) - 2 c_{h_1} t (x_3 + x_4) + 
   b (-1 + x_2 + t (1 - x_2 + x_3) + x_4)\bigr)\\
      \dot{x}_3 &= \frac{1}{2} x_3 \bigl(2 c_{h_2} (-1 + t) (x_2 + x_4) - 2 c_{h_1} t (-1 + x_3 + x_4) + 
   b (x_2 + t (-1 - x_2 + x_3) + x_4)\bigr)\\
  \dot{x}_4 &=  \frac{1}{2} x_4 \bigl(2 c_{h_2} (-1 + t) (-1 + x_2 + x_4) - 2 c_{h_1} t (-1 + x_3 + x_4) + b (-1 + x_2 - t x_2 + t x_3 + x_4)\bigr)
\end{align*}

The system admits four equilibria, of which only one is ever stable: the pure NH state $(x_1^* = 0,\, x_2^* = 1,\, x_3^* = 0,\, x_4^* = 0)$, in which all individuals use strategy N in Task~1 and H in Task~2. This equilibrium is stable for all $t \in [0, 1]$; none of the remaining three equilibria is stable for any value of $t$. The corresponding social welfare is
\begin{equation*}
S(t)=(1-t)(b - c_{h_2})\,,
\end{equation*}
which decreases linearly in $t$.

The welfare difference due to the introduction of genAI is theqrefore
\begin{equation}\label{eq:deltaS-zero-switching-cost}
    \Delta S=\frac{
-2 (c_a - c_{h_2})^2 + \left[ 2 (c_a - c_{h_2})^2 + (b - c_a) m - m^2 \right] t
}{m}\,,
\end{equation}
which increases linearly in $t$, from negative at $t = 0$ (all ``\textit{poetry}'') to positive at $t = 1$ (all ``\textit{busy-work}'').

\end{document}